\documentclass{article}

\usepackage{arxiv}

\usepackage[utf8]{inputenc} 
\usepackage[T1]{fontenc}    
\usepackage{url}            
\usepackage{booktabs}       
\usepackage{amsfonts}       
\usepackage{nicefrac}       
\usepackage{microtype}      
\usepackage{lipsum}		
\usepackage{graphicx}
\usepackage{lineno}
\usepackage[utf8]{inputenc}
\usepackage{lipsum}
\usepackage{multicol}
\usepackage[table]{xcolor}
\usepackage{tikz}
\usepackage{amsmath}
\usepackage{amsthm}
\usepackage{flushend}
\RequirePackage{doi}
\usepackage{lineno,hyperref}
\usepackage[utf8]{inputenc}
\usepackage{lipsum}
\usepackage{multicol}
\usepackage[table]{xcolor}
\usepackage{tikz}
\usepackage{amsmath}
\usepackage{amsthm}
\usepackage{flushend}
\usepackage{multirow}
\usepackage{algorithm}
\usepackage{algpseudocode}
\usepackage{amsmath}
\usepackage{longtable}

\usepackage{lineno}
\usepackage[utf8]{inputenc}
\usepackage{lipsum}
\usepackage{multicol}
\usepackage[table]{xcolor}
\usepackage{tikz}
\usepackage{amsmath}
\usepackage{amsthm}
\usepackage{flushend}
\usepackage{graphicx}
\usepackage{hyperref}
\usepackage{multirow}
\usepackage{longtable}
\usepackage{listings}
\usepackage{float}
\usepackage{tabu}
\usepackage{wasysym}
\usepackage{booktabs}
\usepackage{lscape}

\modulolinenumbers[5]

\newcommand{\RN}[1]{%
	\textup{\uppercase\expandafter{\romannumeral#1}}%
}

\definecolor{bar}{rgb}{0.557,0.663,0.859}
\definecolor{back}{rgb}{0.851,0.882,0.949}

\title{Twenty-two years since revealing cross-site scripting attacks: a systematic mapping and a comprehensive survey}

\author{ {\hspace{1mm}Abdelhakim Hannousse}
	\\
	Department of Computer Science\\
	Universty of 8 May 1945, Guelma\\
	 BP 401, Guelma 24000, Algeria\\
	\texttt{hannousse.abdelhakim@univ-guelma.dz} \\
	\And
	{\hspace{1mm}Salima Yahiouche} \\
	Department of Computer Science\\
	LRS laboratory, Badji Mokhtar University\\
	BP 12, Annaba 23000, Algeria\\
	\texttt{yahiouche.salima@univ-annaba.dz} \\
	\And
	{\hspace{1mm}Mohamed Cherif Nait-Hamoud} \\
	Department of Mathematics and Science Computing\\ 
	Larbi Tebessi University\\
	BP 289, Tebessa, 12000, Algeria\\
	\texttt{mohamed-cherif.nait-hamoud@univ-tebessa.dz} \\
}

\renewcommand{\headeright}{}

\begin{document}
\maketitle

\begin{abstract}
Cross-site scripting (XSS) is one of the major threats menacing the privacy of data and the navigation of trusted web applications. Since its reveal in late 1999 by Microsoft security engineers, several techniques have been developed in the aim to secure web navigation and protect web applications against XSS attacks. The problem became worse with the emergence of advanced web technologies such as Web services and APIs and new programming styles such as AJAX, CSS3 and HTML5. While new technologies enable complex interactions and data exchanges between clients and servers in the network, new programming styles introduce new and complicate injection flaws to web applications. XSS has been and still in the TOP 10 list of web vulnerabilities reported by the Open Web Applications Security Project (OWASP). Consequently, handling XSS attacks became one of the major concerns of several web security communities. In this paper, we contribute by conducting a systematic mapping and a comprehensive survey. We summarize and categorize existent endeavors that aim to protect against XSS attacks and develop XSS-free web applications. The present review covers 147 high quality published studies since 1999 including early publications of 2022. A comprehensive taxonomy is drawn out describing the different techniques used to prevent, detect, protect and defend against XSS attacks. 
Although the diversity of XSS attack types and the scripting languages that can be used to state them, the systematic mapping revealed a remarkable bias toward basic and JavaScript XSS attacks and a dearth of vulnerability repair mechanisms.
The survey highlighted the limitations, discussed the potentials of existing XSS attack defense mechanisms and identified potential gaps. 
\end{abstract}


\section{Introduction}

The use of web applications through the Internet has become an indispensable mean for different business and governmental organizations to reduce costs, speed up activities, improve the quality of services and reach as many targeted people as possible. Users also get immense benefits from online provided services. However, those gains are not without risks; web applications requiring users' registrations through input forms are preferred targets for different hacking attacks, putting their own and users' confidential data at risks. 

Cross-site scripting (XSS) is recognized as one of the most dangerous attacks threatening the security of several types of web applications since 1999. XSS attacks are code injection based attacks; attackers can get benefits from vulnerabilities, also named XSS, discovered in trusted web applications, their related plugins or hosting servers, to inject malicious scripts in the aim to compromise them. This way, explored pages of compromised web applications on users' browsers enable attackers to elevate their access-privileges, reveal sensitive user information retained by browsers such as usernames and passwords on behalf of victim users. 
XSS attacks are mainly caused by the lack of appropriate control mechanisms of user inputs in input forms. This is due to: (1) unfamiliarity of web developers with security issues and practices, (2) business requirements for rapid changes of web services to meet new customers’ needs and (3) the simple nature of the attack itself where user inputs should never be trusted and need to be considered as potential attack vectors. XSS attacks can be launched from client and/or server-sides which makes them difficult to be tracked. The XSS vulnerability appears in of the top 10 list of vulnerabilities reported by the Open Web Application Security Project (OWASP) since 2003 until the last released report in 2021\footnote{The Open Web Application Security Project: \url{https://owasp.org/www-project-top-ten/}}.

The wide prevalence of XSS attacks and the increasing number of endeavors to handle them, motivated us to conduct a systematic mapping study classifying all published studies regarding their focus on tackling the XSS issue since its discovery. We aim to: (1) detect any bias toward specific types of XSS attacks or methodological solutions revealing potential gaps and (2) draw out a comprehensive taxonomy of XSS attacks and their handling techniques from studying contemporary studies. We describe each category of solutions showing their strengths and weaknesses and we compare similar approaches when possible. This helps researchers and practitioners from both academia and industry in keeping track of existing efforts and provides a comprehensive guide to their future research directions.

The sequel of this paper is structured as follows: section~\ref{sec:rw} compares the present review to previously published surveys; section~\ref{sec:sms} details the research methodology adopted for this review; section~\ref{sec:results} presents the results of the search and selection processes; each of the five subsequent sections~\ref{sec:rq1}-\ref{sec:rq5} answers one of the research questions marking the focus of the present study; section~\ref{sec:discussion} summarizes and highlights the results of the review, section~\ref{sec:tv} discusses threats to validity and section~\ref{sec:conclusion} concludes the paper. 

\section{Related reviews}
\label{sec:rw}
Several reviews are found in the literature summarizing existing studies on XSS attacks. In this section, existing surveys are presented chronologically and discussed showing their main focus and findings. 

Malviya el al.~\cite{Malviya2013} classified existing XSS attack defensive solutions into avoidance, prevention, detection and removal. According to the authors, avoidance can be achieved through enforcing input sanitization where prevention can be attained by adopting guidelines from trusted and well-known community projects such as OWASP. Three detection and removal approaches are discussed in the survey: static, dynamic and hybrid approaches. After discussing some techniques from each category, the authors ended up by advocating the use of advanced program analysis, specifically artificial intelligence based solutions that may provide a trade-off between performance, amount of manual work and suitability to  discover new attacks.

Hydara et al.~\cite{Hydara2015} conducted a systematic literature review on XSS. The review included studies published before 2013 focusing on XSS attacks. They identified proposed techniques to handle the XSS issue, their main focus was the identification of most addressed types of XSS attacks and defense techniques. The review revealed much focus on reflected XSS and XSS vulnerabilities detection but less focus on XSS vulnerabilities elimination. Our review is much broader since it covers published papers before and after 2013. It includes complementary research questions and provides a comprehensive survey of existing approaches.

Nithya et al.~\cite{Nithya2015} presented a comprehensive survey of existing approaches on the detection and prevention of XSS attacks. A total of 36 techniques are presented with their limitations and deployment locations. No new taxonomy is proposed. Instead, a traditional classification is adopted: static, dynamic and hybrid.

Deepa and Thilagam~\cite{Deepa2016} presented a broader systematic mapping of defense mechanisms against injection and logic vulnerabilities including SQL injection and XSS flaws. In the mapping, studies are categorized based on the phase of software development life cycle where the defense mechanism is integrated. The authors revealed that there is no single solution to mitigate all the flaws. They advocated much focus on fixing vulnerabilities caused by the source code of web applications.

Gupta et al.~\cite{Gupta2017} proposed a taxonomy for XSS attacks putting much emphasis on XSS worms. State-of-the-art-techniques are qualitatively compared regarding their deployment locations, strengths and weaknesses and handled attacks. The study revealed a lack of appropriate DOM-based attack detection.

Chaudhary et al.~\cite{Chaudhary2018}, Sarmah et al.~\cite{Sarmah2018} and Gupta and Gupta~\cite{Gupta2019} used a  previously known classification based on the location where each solution is installed (client, server, client-server, proxy). Chaudhary et al.~\cite{Chaudhary2018} identified five research gaps deduced from  their qualitative comparison: (1) less attention toward DOM and Mutation XSS attacks, (2) inappropriate discrimination of benign from malicious scripts, (3) improper handling of partial script injection (slight modifications to benign scripts), (4) inappropriate context determination and (5) incomplete sanitization support for new HTML5 features.
Sarmah et al.~\cite{Sarmah2018} discussed existing approaches and tools regarding the area of focus (attack or vulnerability) and detected attacks. Tools are compared regrading their platforms, deployment locations and availability of graphic user interfaces. The authors concluded the study by proposing ten recommendations including putting much emphasis for the detection of unknown variants of XSS; consider HTML5 attacks and handling multi-step attacks caused by XSS. 
Gupta and Gupta~\cite{Gupta2019} adopted a set of qualitative criteria for comparing existing endeavors and ended up with a list of guidelines for designing robust solutions for XSS attacks. The guidelines include developing techniques for nested context determination and sanitization for new HTML5 features.

Liu et al.~\cite{Liu2019} also adopted the traditional classification of existing works on the detection of XSS vulnerabilities (static, dynamic and hybrid). The authors discussed the pros and cons of selected studies together with their abilities to detect different types of XSS attacks. The authors found that no method is perfect and recommended the combination of different techniques at the same time. 

Rodríguez et al.~\cite{Rodriguez2020} summarized tools and methods to mitigate XSS attacks. Selected tools and methods are compared regarding different qualitative attributes including their reliance on artificial intelligence techniques. The study revealed a low trend for the use of machine learning compared to traditional methods. Therefore, they advocated the use of artificial intelligence techniques and investigation of the impact of XSS attacks on Internet of Things and advanced industrial devices.  

Table~\ref{tab:reviews} compares the above discussed surveys regarding five quality attributes: (1) adoption of a systematic search for retrieving papers, (2) details provided for each discussed technique, (3) proposition of an overall taxonomy for attacks and attack defense techniques, (4-5) number and recency of included papers. From the table, 
a single systematic review is conducted by Hydara et al.~\cite{Hydara2015} including papers published before 2013. The systematic mapping conduced by Deepa and Thilagam~\cite{Deepa2016} focused on injection vulnerabilities in general and not only XSS. In this study, we focus only on XSS and we include more and recent publications. Moreover, we present a comprehensive taxonomy that clarifies the scope of each study and classifies, separately, existing techniques used to handle attacks and vulnerabilities. 

	\begin{table}[h]
		\centering
		\small{
			\caption{Comparison of existing reviews:  \CIRCLE: full support, \RIGHTcircle; partial support, \Circle: no support. }
			\label{tab:reviews}       
			\begin{tabular}{llcccrl}
				\hline\noalign{\smallskip}
				study & year & systematic & survey & taxonomy & \# included papers & recency of papers\\
				\noalign{\smallskip}\hline\noalign{\smallskip}
				Malviya et al.~\cite{Malviya2013} & 2013 & \Circle & \CIRCLE & \RIGHTcircle & 18 &  2003-2012\\
				Hydara et al.~\cite{Hydara2015} & 2015 &\CIRCLE  & \Circle & \RIGHTcircle &  115 &  2000-2012\\
				Nithya et al.~\cite{Nithya2015} & 2015 &\Circle & \CIRCLE& \RIGHTcircle  & 36 &  2001-2009\\
				Deepa and Thilagam~\cite{Deepa2016} & 2016 &\CIRCLE & \CIRCLE & \RIGHTcircle & 86 &  2005-2015\\
				Gupta et al.~\cite{Gupta2017} & 2017 & \Circle & \CIRCLE & \RIGHTcircle & 19 &  2006-2015\\
				Chaudhary et al.~\cite{Chaudhary2018} & 2018 & \Circle & \RIGHTcircle & \Circle & 27 &  2006-2015\\
				Sarmah et al.~\cite{Sarmah2018} & 2018 & \Circle & \CIRCLE & \Circle & 65 &  2002-2017\\
				Gupta and Gupta~\cite{Gupta2019} & 2019 & \Circle & \CIRCLE & \Circle & 65 &  2004-2018\\
				Liu et al.~\cite{Liu2019} & 2019 & \Circle & \CIRCLE & \Circle & 30 &  2013-2019\\
				Rodríguez et al.~\cite{Rodriguez2020} & 2020 &\Circle & \CIRCLE & \RIGHTcircle & 67 &  2013-2019\\
				\textbf{This work} & 2022 & \CIRCLE & \CIRCLE & \CIRCLE & 147 &  1999-Early 2022\\
				\noalign{\smallskip}\hline
		\end{tabular}}
	\end{table}

In this section, we present and detail the review protocol adopted for the elaboration of the present study. We followed the reported guidelines of Kitchenham et al.~\cite{Kitchenham2015} and Kuhrmann et al.~\cite{Kuhrmann2017} for the preparation and validation of the protocol. In the subsequent subsections, we describe the set of research questions pointing out the focus of the study, the elaborated process for searching, filtering and selecting relevant papers, and we denote the data extraction and synthesis processes.

\subsection{Research Questions}
\label{sec:questions}
The purpose of this paper is to review existing studies with a primary focus on handling the cross-site scripting issue. This leads to identify current gaps and reveal potential future research directions. Research questions are formulated in light of these aims and the study is conducted with five main questions in mind:

\textbf{RQ1.} How much studies were conducted addressing the XSS issue since its reveal in 1999? This research question is adopted to check whether XSS has gained sufficient attention regarding its prevalence.

\textbf{RQ2.} What type of researches are published addressing the XSS issue? This question aims to identify current research focus on tackling XSS flaws.

\textbf{RQ3.} What type of XSS attacks are addressed by contemporary studies? By this question, we aim to provide a classification of existing XSS attacks and identify the most addressed XSS attack types.

\textbf{RQ4.} What techniques have been used to tackle the XSS issue? This question is considered to provide a global overview of already explored techniques, approaches and technologies handling XSS attacks and vulnerabilities as well as their scopes and applicability levels.

\textbf{RQ5.} What type of evidences are used to validate solution proposals? This question enlightens the type of validation techniques and sources of experimented data if any.

\subsection{Search process}
\label{sec:search}
In order to avoid missing relevant papers addressing the XSS issue, we adopted a generic search string for the automatic search in digital libraries. The adopted search string is constructed from the word \textit{"XSS"} or one of its variations and names. The final adopted search string is :

\begin{small}
	\begin{gather*} 
	\text{"XSS" OR "Cross-site scripting" OR "Cross site scripting"}
	\end{gather*}
\end{small}

We observed that many papers address generic attacks but refer to XSS in their abstracts as a comparable attack or vulnerability. To reduce the number and increase the relevance of retrieved papers, we decided to restrict our search to paper titles. This way, only papers with a main focus on XSS are retrieved. For searching high quality peer-reviewed papers, we performed a primary automatic search within Scopus (\url{https://www.scopus.com}). As suggested by Kuhrmann et al.~\cite{Kuhrmann2017}, we performed a second and individual search within each of the following five online academic libraries:

\begin{enumerate}
	\item IEEE Xplorer (\url{https://ieeexplore.ieee.org})
	\item ACM Digital Library (\url{https://dl.acm.org})
	\item SpringerLink (\url{https://link.springer.com})
	\item ScienceDirect (\url{https://www.sciencedirect.com/})
	\item Wiley Online Library (\url{https://onlinelibrary.wiley.com})
\end{enumerate}

To alleviate the impact of the above taken decision about word search position and to avoid potential missing of relevant studies, we conducted a recursive backward and forward snowballing on approved papers as suggested by Wohllin~\cite{Wohlin2016}. Therefore, references and citations of approved papers are screened for relevance regardless of the presence of search strings in their titles. The snowballing is repeated for each new approved paper. Google Scholar is used for exploring citing papers. 

\subsection{Study selection process}
The list of found papers was subject to two separate screening stages. Firstly, the metadata including titles and abstracts are read and checked for relevance and whether they meet a subset of related inclusion/exclusion criteria. Secondly, full texts of papers are examined to check if they meet the rest of inclusion/exclusion criteria and achieve a reasonable quality assessment score.  Metadata screening is performed separately by the two first authors; decisions are exchanged and conflicts are discussed and solved.

\subsubsection{Inclusion and exclusion criteria}
A set of inclusion and exclusion criteria are considered for the selection of appropriate and relevant papers. In this study, only English published journal and conference papers with available full texts are selected. The search process is conducted in October 2021; to avoid missing interesting papers, we searched papers published since 1999, which is the year where the first XSS issue was reported by Microsoft security engineers, we also considered early publications for complete search. The entire list of inclusion and exclusion criteria adopted for the selection process, in this study, are described in Table~\ref{tab:inclExcl}.

	\begin{table}[h]
		\centering
		\small{
			\caption{Inclusion and Exclusion criteria: I$_{i}$ : inclusion criterion, E$_{i}$: exclusion criterion}
			\label{tab:inclExcl}       
			\begin{tabular}{lp{15cm}}
				\hline\noalign{\smallskip}
				ID & Criteria\\
				\noalign{\smallskip}\hline\noalign{\smallskip}
				I1 & papers addressing any aspect regarding XSS attacks and/or vulnerabilities\\
				I2 & papers written in English\\
				I3 & peer-reviewed journal or conference papers\\
				I4 & papers with full text available\\
				I5 & early publications\\
				\noalign{\smallskip}\hline\noalign{\smallskip}
				E1 & papers not addressing XSS such as general-purpose attacks or vulnerabilities detection tools or methods.\\ 
				E2 & duplicate papers (older versions of already retrieved papers) and short papers with the presence of extended versions. Snowballing is performed for old and new versions to avoid missing relevant papers.\\
				E3 & white papers, technical reports, reviews/surveys, patents, feature articles, very short papers (less than 3 pages length), books and book chapters.\\
				\noalign{\smallskip}\hline
		\end{tabular}}
	\end{table}

\subsubsection{Quality assessment}
\label{sec:qa}
For including only high quality studies, all the selected papers from the previous stage were subject to a strict quality assessment check. For this sake, a checklist of 5 items is used to assess the quality of all found studies; only those with a total score greater or equal 4 are considered. This stage is performed separately by the two first authors and conflicts are resolved by discussion. Table~\ref{tab:qa} describes the adopted criteria together with the range of values associated to each item. QA1 promotes studies where the main focus is XSS against others considering XSS with similar issues such as SQL injection. Therefore, a complete score value is associated only to papers addressing only the XSS issue. QA2 evaluates the clarity and the adopted scheme or methodology of each study. Some papers are not self-contained and does not provide sufficient details on the proposed solution or the research methodology, those papers are either given 0.5 or 0 score value according to the provided details given in the paper. QA3 checks the appropriateness of the evaluation process adopted for the proposed solution. A high value for QA3 is given to papers that uses a reasonable size of test cases or datasets, a comparison with similar works and uses appropriate evaluation metrics. QA4 checks the popularity of the publication venue; paper sources are ranked following the CORE\footnote{CORE Rankings Portal:~\url{http://portal.core.edu.au/conf-ranks/}} conference ranking: A or B (+1), C (+0.5) and not ranked (0) and Scientific Journal Rankings Portal\footnote{Scientific Journal Rankings - SCImago:~\url{https://www.scimagojr.com/}} (SJR): Q1-2 (+1), Q3-4 (0.5), not ranked (0).  Finally, QA5 evaluates the impact and influence of the study on the research field; consequently, score value 1 is attributed to papers with number of citations greater or equal to the number of years since the paper was published and 0 otherwise. This is adopted to avoid penalizing early publications. Google Scholar is used to identify the number of citations associated to each paper.

	\begin{table}[h]
		\centering
		\small{
			\caption{Quality assessment criteria}
			\label{tab:qa}       
			\begin{tabular}{llr}
				\hline\noalign{\smallskip}
				ID & Criteria & Value\\
				\noalign{\smallskip}\hline\noalign{\smallskip}
				QA1 & Does the study exclusively focus on XSS attacks and vulnerabilities? & Yes(1)/No(0.5)\\
				QA2 & Does the study present a clear solution or adequate research methodology? & $\{$0, 0.5, 1$\}$\\
				QA3 & Does the study used appropriate methods to validate the proposed solution?  & $\{$0, 0.5, 1$\}$\\
				QA4 & Has the study been published in a well-established source (journal or conference)?  & $\{$0, 0.5, 1$\}$\\
				QA5 & Has the study been fairly cited by other papers? & $\{$0, 1$\}$\\
				\noalign{\smallskip}\hline
		\end{tabular}}
	\end{table}

\subsection{Data extraction process}
Data are extracted manually from selected and approved papers by the first two authors and validated by the third author. Disagreements are resolved by discussion to reach a consensus. 
In this phase, we followed the guidelines of Peterson et al.~\cite{Petersen2008} where a data extraction form is designed as illustrated in Table~\ref{tab:dataextraction}. Each selected paper is thoroughly analyzed by the authors where all the required data for answering the research questions presented in section~\ref{sec:questions} are extracted and detailed in the form. Besides the metadata used for answering RQ1, more detailed information required for answering the other research questions are made explicit in the designed form. Those information include: type of study, detailed description of the proposed solution or analysis process, applicability level of the proposal if any, techniques and lists of data sources used for validation.

	\begin{table}[h]
		\centering
		\small{
			\caption{Data extraction form}
			\label{tab:dataextraction}       
			\begin{tabular}{llll}
				\hline\noalign{\smallskip}
				ID & Data item & Description&RQ\\
				\noalign{\smallskip}\hline\noalign{\smallskip}
				D1& Year  & year of the publication& RQ1\\
				D2& Source & venue of the publication& RQ1\\
				D3& Type of publication & conference or journal paper& RQ1\\ 
				D4& Type of the study & analysis, solution proposal & RQ2\\
				D5& Focus & attacks, vulnerabilities & RQ2-4\\
				D6&  Description & a short overview of the proposed solution & RQ2-4\\
				D7 & Taxonomy & proposed or adopted classification or taxonomy, if any & RQ3\\
				D8&  Types of XSS &  type of XSS attack(s) addressed by the study & RQ3\\
				D9&  Detailed focus of the study & prevention, detection, protection, etc. & RQ4\\
				D10 & Applicability & applicability level of the solution: client, server, proxy, etc. & RQ4\\
				D11&  Evaluation & validation process adopted by the study& RQ5\\
				D12& Sources & list of datasets or sources used for validation if any. & RQ5\\  
				\noalign{\smallskip}\hline
		\end{tabular}}
	\end{table}

\subsection{Data synthesis}
After data extraction from approved papers, descriptive statistics and frequency analysis are performed to answer research questions RQ1, RQ2 and RQ5. The results are illustrated using tables and visualized using appropriate plots and graphics. In addition, a thematic analysis is performed based on the guidelines suggested by Cruzes et al.~\cite{Cruzes2011} to answer RQ3 and RQ4. Accordingly, patterns in the data are firstly identified and then grouped into distinct themes; those themes are reviewed and refined to produce final themes. The three authors are involved for the elaboration of this task.

\section{Results of the search and selection processes}
\label{sec:results}
\begin{figure}[h]
	\centering
	\includegraphics[width=.65\textwidth]{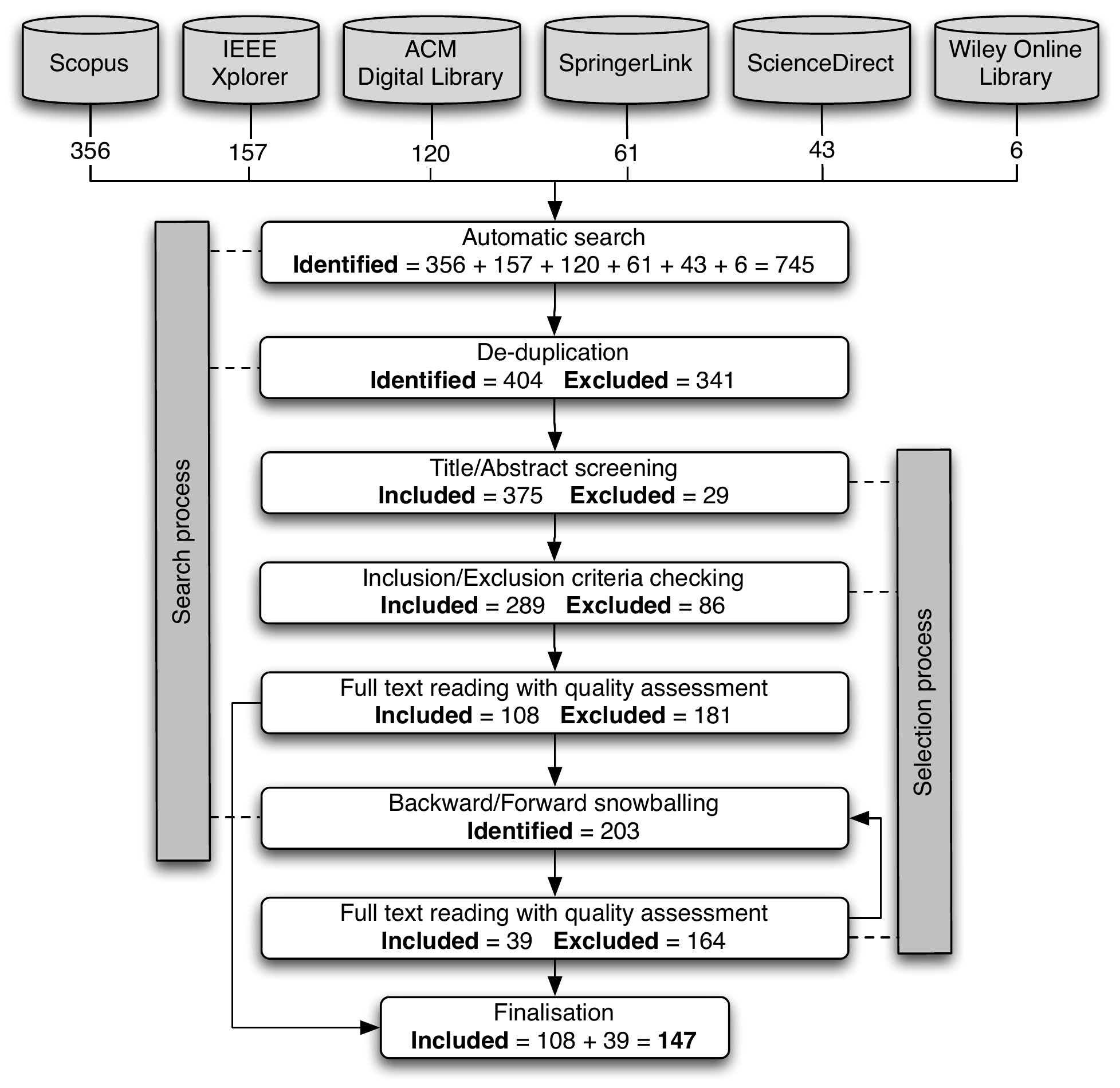}
	\caption{Results of the search and selection processes}
	\label{fig:search}
\end{figure}

The search process started by exploring 6 different academic search engines using the search query described in section~\ref{sec:search}; this step yielded 745 papers; after de-duplication, 404 distinct papers remained. Identified papers from the automatic search were subject for eligibility check. The title and abstract screening revealed that 29 papers were irrelevant to XSS and 86 more papers did not meet our inclusion and exclusion criteria. From the 289 remaining papers, only 108 papers reached the threshold value adopted for the quality assessment check (i.e.; have quality scores greater or equal 4). Those papers formed the initial set of approved papers. The recursive backward and forward snowballing over the 108 approved papers identified 203 new papers but only 39 of them passed the eligibility check which finally resulted in a total of 147 papers. Figure~\ref{fig:search} summarizes the overall steps of search and selection processes and indicates intermediate search results including the number of identified, included and excluded papers in each step. 

\section{How much studies were conducted addressing the XSS issue since its reveal in 1999?}
\label{sec:rq1}
\begin{figure}[h]
	\centering
	\includegraphics[width=.9\textwidth]{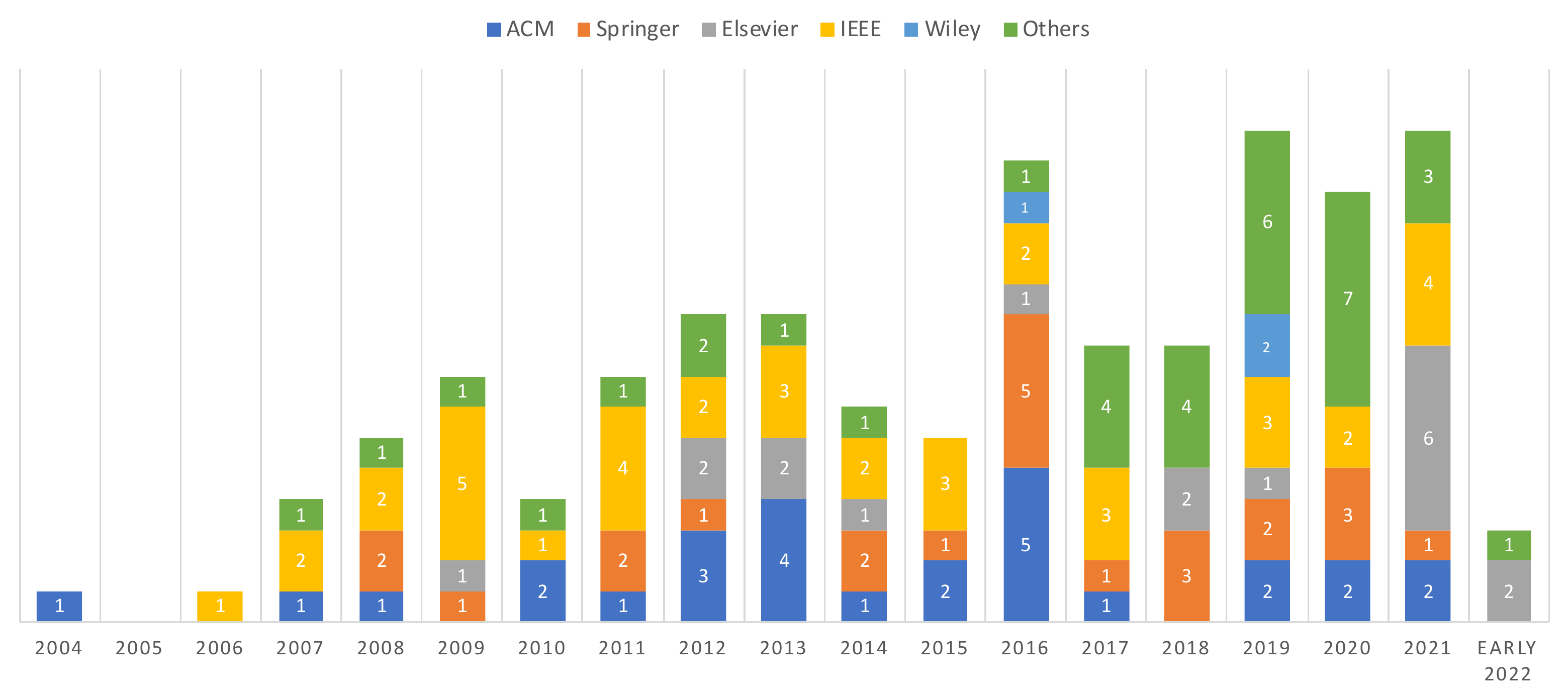}
	\caption{Distribution of included studies over years and publishers.}
	\label{fig:peryear}    
\end{figure}

The adopted search and selection processes resulted in the retrieval of 147 distinct, relevant and high quality papers. Figure~\ref{fig:peryear} shows the distribution of included papers according to their publication years and publishers. Remarkably, although the early detection of the XSS issue in 1999, the attack started earning academic research attentions since 2016 and higher attentions were given in the last years (2019 and 2021). 

Table~\ref{tab:studies} shows the entire list of included studies with their publication years and their correspondent detailed quality assessment scores. The studies in the table are ordered descendingly following their publication years. It can be noticed that most papers (97/147) are conference papers and only 50 from the entire set are journal publications. This is due to the publication speed of conference papers compared with journals. Only 36 papers have a total quality assessment score where 52 papers have the exact threshold value for eligibility. Specifically, 32 papers have been published in low ranked journals or conferences and 8 of them are penalized since they address other issues in addition to XSS, the others do not completely satisfy QA2 and QA3 (i.e.; do not present a clear solution and have lacks in their adopted validation methods).

	\begin{longtable}{lccrrrrrr}
		\caption{List of included studies with their detailed quality assessment scores: J:~Journal paper, C:~Conference paper}
		\label{tab:studies} \\      
		\hline\noalign{\smallskip}
		Ref. & Year & Type & QA1 & QA2 & QA3 & QA4 & QA5 & Total QA Score\\
		\noalign{\smallskip}\hline\noalign{\smallskip}
		Huang et al.~\cite{Huang2004}& 2004& C & 0.5 & 1 & 0.5 & 1 & 1 & 4 \\
		Jovanovic et al.~\cite{Jovanovic2006a}&2006& C & 1 &1 & 0.5 & 1 & 1 & 4.5\\
		Shanmugam and Ponnavaikko~\cite{Shanmugam2007a}&2007& C & 1 & 1 & 0.5 & 0.5 & 1 & 4\\
		Shanmugam and Ponnavaikko~\cite{Shanmugam2007b}&2007& C & 1 & 1 & 0.5 & 0.5 & 1 & 4\\
		Vogt et al.~\cite{Vogt2007}&2007& C & 1 & 0.5 & 1 & 1 & 1 & 4.5\\
		Jim et al.~\cite{Jim2007}&2007& C & 0.5 & 1 & 0.5 & 1 & 1 & 4\\
		Martin and Lam~\cite{Martin2008}&2008& C & 0.5 & 1 & 0.5 & 1 & 1 & 4\\
		McAllister et al.~\cite{McAllister2008}&2008& C & 1 & 0.5 & 0.5 & 1 & 1 & 4\\
		Balzarotti et al.~\cite{Balzarotti2008}&2008& C & 0.5 & 1 & 0.5 & 1 & 1 & 4\\
		Wassermann and Su~\cite{Wassermann2008}&2008& C & 1 & 1 & 0.5 & 1 & 1 & 4.5\\
		Johns et al.~\cite{Johns2008}&2008& C & 1 & 1 & 0.5 & 1 & 1 & 4.5\\
		Bisht and Venkatakrishnan~\cite{Bisht2008}&2008& C & 1 & 1 & 0.5 & 0.5 & 1 & 4\\
		Iha and Doi~\cite{Iha2009}&2009& C & 1 & 1 & 0.5 & 1 & 1 & 4.5\\
		Kiezun et al.~\cite{Kiezun2009}&2009& C & 0.5 & 1 & 0.5 & 1 & 1 & 4\\
		Louw and Venkatakrishnan~\cite{Louw2009}&2009& C & 1 & 1 & 1 & 1 & 1 & 5\\
		Kirda et al.~\cite{Kirda2009}&2009& J & 1 & 1 & 0.5 & 1 & 1 & 4.5\\
		Sun et al.~\cite{Sun2009}&2009& C & 1 & 1 & 0.5 & 1 & 1 & 4.5\\
		Nadji et al.~\cite{Nadji2009}&2009& C & 1 & 0.5 & 1 & 1 & 1 & 4.5\\
		Barth et al.~\cite{Barth2009} &2009& C & 1 & 1 & 0.5 & 1 & 1 & 4.5\\
		Wurzinger et al.~\cite{Wurzinger2009}&2009& C & 1 & 1 & 0.5 & 1 & 1 & 4.5\\
		Gebre et al.~\cite{Gebre2010}&2010& C & 1 & 1 & 0.5 & 0.5 & 1 & 4\\
		Bates et al.~\cite{Bates2010}&2010& C & 1 & 1 & 0.5 & 1 & 1 & 4.5\\
		Stamm et al.~\cite{Stamm2010}&2010& C & 0.5 & 1 & 0.5 & 1 & 1 & 4\\
		Athanasopoulos et al.~\cite{Athanasopoulos2010}&2010& C & 1 & 1 & 1 & 0 & 1 & 4\\
		Weinberger et al.~\cite{Weinberger2011}&2011& C & 1 & 1 & 1 & 1 & 1 & 5\\
		Samuel et al.~\cite{Samuel2011}&2011& C & 1 & 1 & 0.5 & 1 & 1 & 4.5\\
		Hooimeijer et al.~\cite{Hooimeijer2011}&2011& C & 1 & 1 & 0.5 & 1 & 1 & 4.5\\
		Mui and Frankl~\cite{Mui2011}&2011& C & 0.5 & 1 & 0.5 & 1 & 1 & 4\\
		Wang et al.~\cite{Wang2011}&2011& C & 1 & 1 & 0.5 & 0.5 & 1 & 4\\
		Shariar and Zulkernine~\cite{Shahriar2011}&2011& C & 1 & 1 & 0.5 & 0.5 & 1 & 4\\
		Avancini and Ceccato~\cite{Avancini2011}&2011& C & 1 & 1 & 0.5 & 0.5 & 1 & 4\\
		Barua et al.~\cite{Barua2011}&2011& C & 1 & 1 & 1 & 1 & 1 & 5\\
		Scholte et al.~\cite{Scholte2012a}&2012& C & 0.5 & 1 & 1 & 1 & 1 & 4.5\\
		Shar and Tan~\cite{Shar2012}&2012& C & 1 & 1 & 0.5 & 0.5 & 1 & 4\\
		Shar and Tan~\cite{Shar2012a}&2012& J & 1 & 1 & 1 & 1 & 1 & 5\\
		Nunan et al.~\cite{Nunan2012}&2012& C & 1 & 1 & 1 & 1 & 1 & 5\\
		Van Acker et al.~\cite{Van-Acker2012}&2012& C & 1 & 1 & 1 & 1 & 1 & 5\\
		Van-Gundy and Chen~\cite{Van-Gundy2012}&2012& J & 1 & 1 & 1 & 1 & 1 & 5\\
		Cao et al.~\cite{Cao2012}&2012& C & 1 & 1 & 0.5 & 1 & 1 & 4.5\\
		Scholte et al.~\cite{Scholte2012}&2012& C & 0.5 & 1 & 0.5 & 1 & 1 & 4\\
		Pelizzi and Sekar~\cite{Pelizzi2012}&2012& C & 1 & 1 & 1 & 1 & 1 & 5\\
		Sundareswaran and Squicciarini~\cite{Sundareswaran2012}&2012& C & 1 & 1 & 0.5 & 1 & 1 & 4.5\\
		Lekies et al.~\cite{Lekies2013}&2013& C & 1 & 1 & 1 & 1 & 1 & 5\\
		Faghani and Nguyen~\cite{Faghani2013}&2013& J & 1 & 1 & 1 & 1 & 1 & 5\\
		Avancini and Ceccato~\cite{Avancini2013a}&2013& C & 1 & 1 & 0.5 & 1 & 1 & 4.5\\
		Avancini and Ceccato~\cite{Avancini2013}&2013& J & 1 & 1 & 1 & 1 & 1 & 5\\
		Doupé et al.~\cite{Doupe2013}&2013& C & 1 & 1 & 1 & 1 & 1 & 5\\
		Das et al.~\cite{Das2013}&2013& J & 1 & 0.5 & 0.5 & 1 & 1 & 4\\
		Tripp et al.~\cite{Tripp2013}&2013& C & 1 & 1 & 0.5 & 1 & 1 & 4.5\\
		Duchène et al.~\cite{Duchene2013}&2013& C & 1 & 1 & 1 & 1 & 1 & 5\\
		Heiderich et al.~\cite{Heiderich2013}&2013& C & 1 & 1 & 0.5 & 1 & 1 & 4.5\\
		Shar and Tan~\cite{Shar2013}&2013& J & 0.5 & 1 & 1 & 1 & 1 & 4.5\\
		Vernotte et al.~\cite{Vernotte2014}&2014& C & 1 & 0.5 & 0.5 & 1 & 1 & 4\\
		Rocha and Souto~\cite{Rocha2014}&2014& C & 1 & 0.5 & 0.5 & 1 & 1 & 4\\
		Duchène et al.~\cite{Duchene2014}&2014& C & 1 & 1 & 1 & 0 & 1 & 4\\
		Wang et al.~\cite{Wang2014} &2014& C & 1 & 0.5 & 0.5 & 1 & 1 & 4\\
		Stock et al.~\cite{Stock2014}&2014& C & 1 & 1 & 1 & 1 & 1 & 5\\
		Johns~\cite{Johns2014}&2014& J & 1 & 1 & 0.5 & 0.5 & 1 & 4\\
		Javed and Schwenk~\cite{Javed2014a}&2014& C & 1 & 1 & 0.5 & 0.5 & 1 & 4\\
		Bozic et al.~\cite{Bozic2015a}&2015& C & 1 & 1 & 1 & 1 & 1 & 5\\
		Fazzini et al.~\cite{Fazzini2015}&2015& C & 1 & 1 & 0.5 & 1 & 1 & 4.5\\
		Parameshwaran et al.~\cite{Parameshwaran2015}&2015& C & 1 & 1 & 0.5 & 1 & 1 & 4.5\\
		Bozic et al.~\cite{Bozic2015c}&2015& C & 1 & 1 & 1 & 1 & 1 & 5\\
		Stock et al.~\cite{Stock2015}&2015& C & 1 & 1 & 1 & 1 & 1 & 5\\
		Zhang et al.~\cite{Zhang2015}&2015& J & 1 & 1 & 0.5 & 1 & 1 & 4.5\\
		Simos and Kleine~\cite{Simos2016}&2016& C & 1 & 1 & 0.5 & 0.5 & 1 & 4\\
		Gupta and Gupta~\cite{Gupta2016b1}&2016& C & 1 & 0.5 & 0.5 & 1 & 1 & 4\\
		Weichselbaum et al.~\cite{Weichselbaum2016}&2016& C & 1 & 1 & 1 & 1 & 1 & 5\\
		Pan et al.~\cite{Pan2016}&2016& C & 1 & 1 & 1 & 1 & 1 & 5\\
		Pan and Mao~\cite{Pan2016a}&2016& C & 1 & 1 & 0.5 & 1 & 1 & 4.5\\
		Mitropoulos et al.~\cite{Mitropoulos2016}&2016& J & 1 & 1 & 0.5 & 1 & 1 & 4.5\\
		Steinhauser and Gauthier~\cite{Steinhauser2016}&2016& C & 1 & 1 & 1 & 0 & 1 & 4\\
		Ahmed and Ali~\cite{Ahmed2016}&2016& J & 1 & 1 & 0.5 & 1 & 1 & 4.5\\
		Lin and Barcelo~\cite{Lin2016}&2016& C & 1 & 1 & 1 & 1 & 1 & 5\\
		Yan and Qiao~\cite{Yan2016}&2016& C & 1 & 1 & 0.5 & 0.5 & 1 & 4\\
		Pan et al.~\cite{Pan2016b}&2016& J & 1 & 1 & 1 & 1 & 0 & 4\\
		Gupta et al~\cite{Gupta2016c}&2016& C & 1 & 1 & 1 & 0 & 1 & 4\\
		Bazzoli et al.~\cite{Bazzoli2016}&2016& C & 1 & 1 & 1 & 0 & 1 & 4\\
		Gupta and Gupta~\cite{Gupta2016e}&2016& J & 1 & 1 & 0.5 & 1 & 1 & 4.5\\
		Gupta and Gupta~\cite{Gupta2016}&2016& J & 1 & 1 & 0.5 & 1 & 1 & 4.5\\
		Goswami et al.~\cite{Goswami2017}&2017& J & 1 & 0.5 & 0.5 & 1 & 1 & 4\\
		Lekies et al.~\cite{Lekies2017}&2017& C & 1 & 1 & 1 & 1 & 1 & 5\\
		Mohammadi et al.~\cite{Mohammadi2017}&2017& C & 1 & 1 & 0.5 & 1 & 1 & 4.5\\
		Pan and Mao~\cite{Pan2017}&2017& C & 1 & 1 & 0.5 & 1 & 1 & 4.5\\
		Marashdih and Zaaba~\cite{Marashdih2017}&2017& C & 1 & 1 & 0.5 & 0.5 & 1 & 4\\
		Heiderich et al.~\cite{Heiderich2017}&2017& C & 1 & 1 & 0.5 & 1 & 1 & 4.5\\
		Gupta and Gupta~\cite{Gupta2017a}&2017& J & 1 & 1 & 0.5 & 0.5 & 1 & 4\\
		Marashdih et al.~\cite{Marashdih2017a}&2017 & J & 1 & 1 & 0.5 & 0.5 & 1 & 4\\
		Rathore et al.~\cite{Rathore2017} &2017& J & 1 & 1 & 0.5 & 0.5 & 1 & 4\\
		Ayeni et al.~\cite{Ayeni2018}&2018& J & 1 & 1 & 0.5 & 1 & 1 & 4.5\\
		Gupta et al.~\cite{Gupta2018d}&2018& J & 1 & 1 & 1 & 1 & 1 & 5\\
		Mereani and  Howe~\cite{Mereani2018a}&2018& C & 1 & 1 & 1 & 0.5 & 1 & 4.5\\
		Melicher et al.~\cite{Melicher2018}&2018& C & 1 & 1 & 1 & 1 & 1 & 5\\
		Gupta and Gupta~\cite{Gupta2018c}&2018& J & 1 & 1 & 1 & 0.5 & 1 & 4.5\\
		Wang et al.~\cite{Wang2018}&2018& J & 1 & 0.5 & 1 & 1 & 1 & 4.5\\
		Yamazaki et al.~\cite{Yamazaki2018}&2018& C & 1 & 1 & 1 & 1 & 0 & 4\\
		Lalia and Sarah~\cite{Lalia2018}&2018& C & 1 & 1 & 0.5 & 0.5 & 1 & 4\\
		Gupta and Gupta~\cite{Gupta2018e}&2018& J & 1 & 1 & 1 & 1 & 1 & 5\\
		Gupta et al.~\cite{Gupta2019a}&2019& J & 1 & 1 & 0.5 & 0.5 & 1 & 4\\
		Garn et al.~\cite{Garn2019}&2019& C & 1 & 0.5 & 0.5 & 1 & 1 & 4\\
		Chaudhary et al.~\cite{Chaudhary2019}&2019& J & 1 & 1 & 0.5 & 0.5 & 1 & 4\\
		Lv et al.~\cite{Lv2019}&2019& C & 1 & 1 & 0.5 & 1 & 1 & 4.5\\
		Zhou and Wang~\cite{Zhou2019}&2019& J & 1 & 1 & 1 & 1 & 1 & 5\\
		Zhang et al.~\cite{Zhang2019}&2019& C & 1 & 1 & 0.5 & 1 & 1 & 4.5\\
		Steinhauser and Tuma~\cite{Steinhauser2019}&2019& J & 1 & 1 & 1 & 1 & 1 & 5\\
		Steffens et al.~\cite{Steffens2019}&2019& C & 1 & 1 & 1 & 1 & 1 & 5\\
		Wijayarathna and Arachchilage~\cite{Wijayarathna2019}&2019& C & 1 & 1 & 0.5 & 1 & 1 & 4.5\\
		Chaliasos et al.~\cite{Chaliasos2019}&2019& C & 1 & 1 & 0.5 & 1 & 1 & 4.5\\
		Mokbal et al.~\cite{Mokbal2019}&2019& J & 1 & 1 & 1 & 1 & 1 & 5\\
		Iqbal et al.~\cite{Iqbal2019}&2019& C & 1 & 1 & 1 & 1 & 0 & 4\\
		Simos et al.~\cite{Simos2019}&2019& C & 1 & 1 & 0.5 & 0.5 & 1 & 4\\
		Fang et al.~\cite{Fang2019}&2019& J & 1 & 1 & 0.5 & 1 & 1 & 4.5\\
		Musch et al.~\cite{Musch2019}&2019& C & 1 & 1 & 0.5 & 1 & 1 & 4.5\\
		Niakanlahiji and Jafarian~\cite{Niakanlahiji2019}&2019& J & 1 & 1 & 0.5 & 1 & 1 & 4.5\\
		Kadhim and Gaata~\cite{Kadhim2020}&2020& J & 1 & 1 & 0.5 & 0.5 & 1 & 4\\
		Zhang et al.~\cite{Zhang2020} &2020& J & 1 & 1 & 1 & 1 & 1 & 5\\
		Li et al.~\cite{Li2020a}&2020& J & 1 & 1 & 0.5 & 1 & 1 & 4.5\\
		Mokbal et al.~\cite{Mokbal2020}&2020& J & 1 & 1 & 1 & 1 & 1 & 5\\
		Steinhauser and Tuma~\cite{Steinhauser2020}&2020& J & 1 & 1 & 1 & 0 & 1 & 4\\
		Gupta et al.~\cite{Gupta2020a}&2020& J & 1 & 1 & 0.5 & 1 & 1 & 4.5\\
		Schuckert et al.~\cite{Schuckert2020}&2020& C & 1 & 1 & 0.5 & 1 & 1 & 4.5\\
		Nagarjun and Ahamad~\cite{Nagarjun2020}&2020& J & 1 & 1 & 1 & 0.5 & 1 & 4.5\\
		Xu et al.~\cite{Xu2020}&2020& J & 1 & 1 & 1 & 1 & 1 & 5\\
		Fang et al.~\cite{Fang2020}&2020& J & 1 & 1 & 0.5 & 1 & 1 & 4.5\\
		Buyukkayhan et al.~\cite{Buyukkayhan2020}&2020& C & 1 & 1 & 0.5 & 1 & 1 & 4.5\\
		Li et al.~\cite{Li2020}&2020& C & 1 & 1 & 0.5 & 0.5 & 1 & 4\\
		Bui et al.~\cite{Bui2020}&2020& C & 1 & 1 & 0.5 & 1 & 1 & 4.5\\
		Chaudhary et al.~\cite{Chaudhary2020}&2020& C & 1 & 1 & 1 & 0 & 1 & 4\\
		Talib and Doh~\cite{Talib2021}&2021& J & 1 & 1 & 1 & 0.5 & 1 & 4.5\\
		Eriksson et al.~\cite{Eriksson2021}&2021& C & 1 & 0.5 & 1 & 1 & 1 & 4.5\\
		Garn et al.~\cite{Garn2021}&2021& C & 1 & 1 & 1 & 1 & 1 & 5\\
		Malviya et al.~\cite{Malviya2021}&2021& J & 1 & 1 & 1 & 1 & 1 & 5\\
		Caturano~ et al.~\cite{Caturano2021}&2021& J & 1 & 1 & 0.5 & 1 & 1 & 4.5\\
		Chaudhary et al.~\cite{Chaudhary2021}&2021& J & 1 & 1 & 1 & 1 & 1 & 5\\
		Frempong et al.~\cite{Frempong2021}&2021& C & 1 & 0.5 & 0.5 & 1 & 1 & 4\\
		Leithner et al.~\cite{Leithner2021}&2021& J & 1 & 1 & 0.5 & 1 & 1 & 4.5\\
		Wang et al.~\cite{Wang2021}&2021& C & 1 & 1 & 0.5 & 1 & 1 & 4.5\\
		Tariq et al.~\cite{Tariq2021}&2021& J & 1 & 1 & 1 & 1 & 1 & 5\\
		Mereani and Howe~\cite{Mereani2021}&2021& C & 1 & 1 & 0.5 & 0.5 & 1 & 4\\
		Maurel et al.~\cite{Maurel2021}&2021& C & 1 & 1 & 0.5 & 1 & 1 & 4.5 \\
		Bensalim et al.~\cite{Bensalim2021}&2021& C & 1 & 1 & 1 & 0 & 1 & 4\\
		Melicher et al.~\cite{Melicher2021}&2021& C & 1 & 1 & 1 & 0 & 1 & 4\\
		Mokbal et al.~\cite{Mokbal2021}&2021& J & 1 & 1 & 1 & 1 & 1 & 5\\
		Pazos et al.~\cite{Pazos2021}&2021& C & 1 & 1 & 0.5 & 1 & 1 & 4.5\\
		Wang et al.~\cite{Wang2022}&2022& J & 1 & 0.5 & 1 & 1 & 1 & 4.5\\
		Mokbal et al.~\cite{Mokbal2022}&2022& J & 1 & 1 & 0.5 & 1 & 1 & 4.5\\
		Liu et al.~\cite{Liu2022}&2022& J & 1 & 1 & 0.5 & 1 & 1 & 4.5\\
		
		\noalign{\smallskip}\hline
		
	\end{longtable}

\begin{figure}[h]
	\centering
	\includegraphics[width=.55\textwidth]{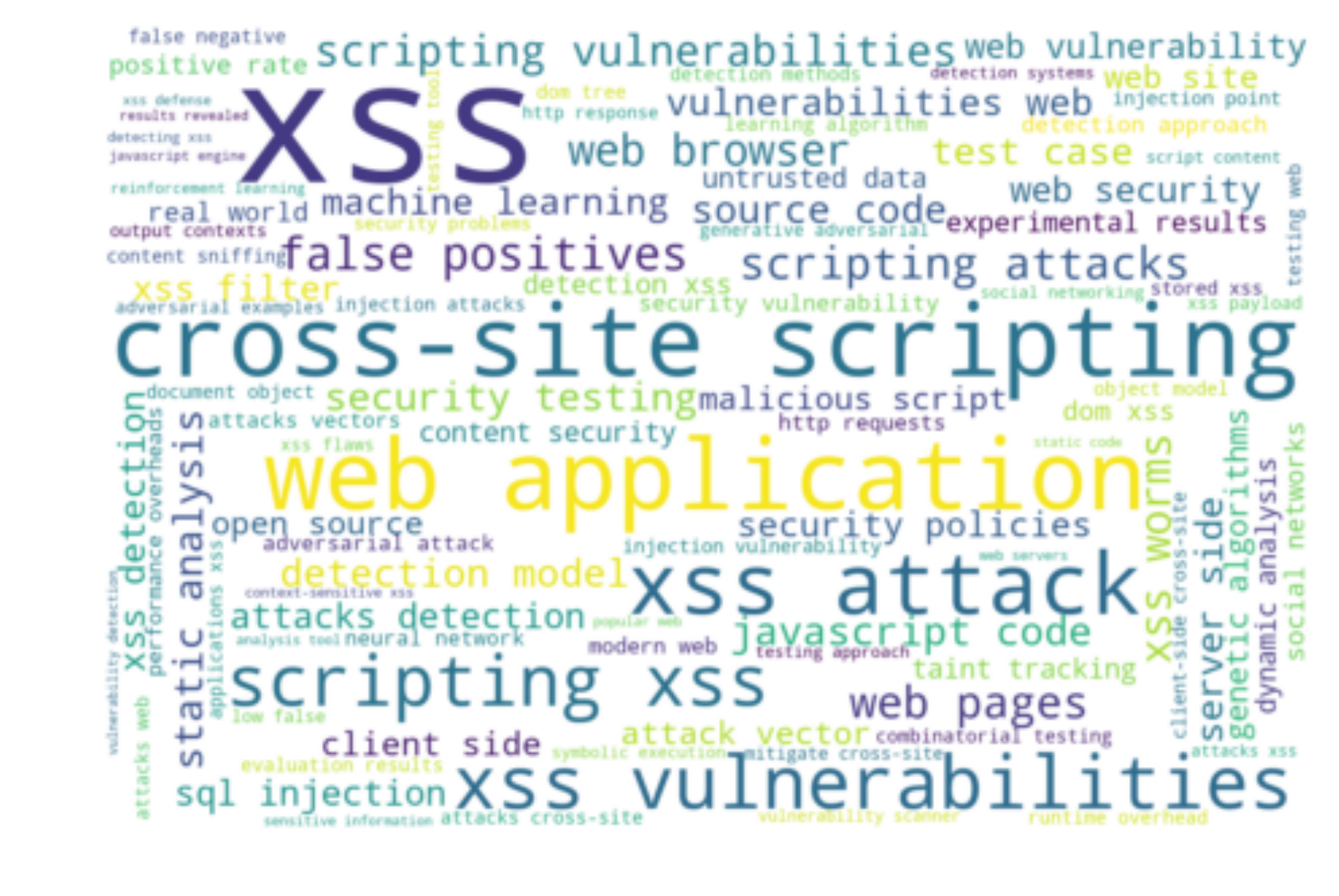}
	\caption{Word cloud constructed using titles and abstracts of included studies}
	\label{fig:wordcloud}
\end{figure}

As suggested by Kuhrmann et al.~\cite{Kuhrmann2017}, a Word Cloud of frequent terms used in titles and abstracts of included papers is shown in Figure~\ref{fig:wordcloud}. Terms are firstly extracted using 1-2grams and then filtered to keep only significant ones. For preserving the visibility of the figure, only top 100 terms are kept. Figure~\ref{fig:wordcloud} illustrates that \textit{XSS}, \textit{cross-site scripting} and \textit{web application} are the most frequent terms, which is consistent with the adopted terms in the search query. Moreover, the figure illustrates a similar interest on XSS attacks and vulnerabilities as well as static and dynamic analysis and client and server side (have quite similar size in the cloud). Moreover, the word \textit{detection} appears frequently in conjunction with terms such as \textit{model}, \textit{attacks} and \textit{XSS}; this indicates that much focus is given to the detection of the problem compared with other techniques such as protection. However, the validation of these preliminary observations requires more exhaustive analysis which is the focus of the proposed research questions steering the present review.

\section{What type of researches are published addressing the XSS issue? (RQ2)}
\label{sec:rq2}
\begin{figure}[h]
	\centering
	\includegraphics[width=.9\textwidth]{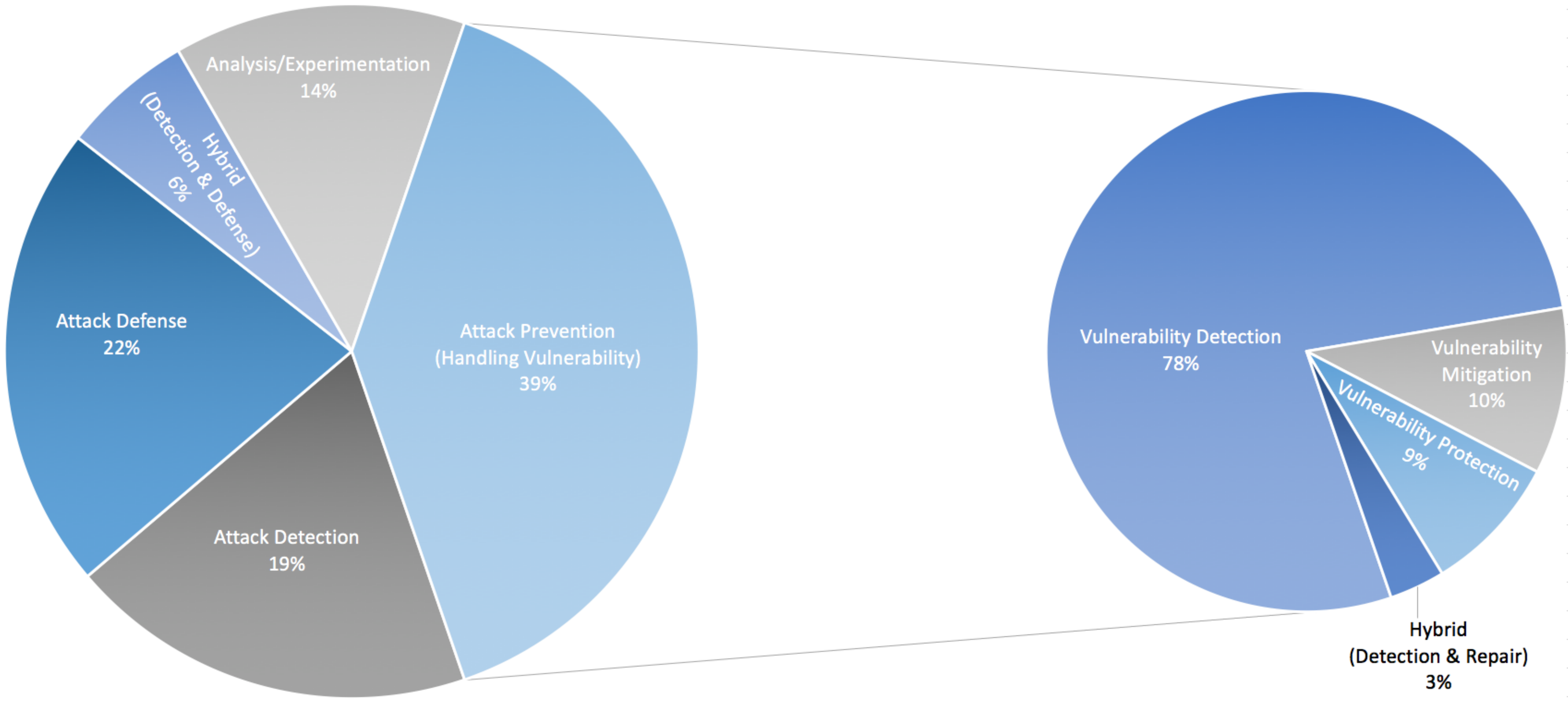}
	\caption{Distribution of included studies per research focus}
	\label{fig:categories}
\end{figure}

Identified studies can be classified into two main categories: \textit{analysis/experimentation} and \textit{solution proposals}. The focus of studies from the former category is one of three kinds: (1) examine the prevalence of the attack or one of its variants in the wild, (2) identify the attack underlying features in the aim to provide proper countermeasures and (3) compare existing solutions highlighting their achievements and unfulfillments. Studies of the latter category can also be classified according to their contribution to handle XSS attacks into: \textit{prevention}, \textit{detection}, \textit{defense} and \textit{hybrid}. Attack detection based solutions focus on how to distinguish attacks from normal behaviors on already deployed and active web applications. Detected attacks are generally reported to web application administrators or victims to take appropriate actions. Attack defense based solutions define a set of countermeasures taken automatically to thwart XSS attacks when they occur. They either provide means to defend against already known attacks or attack vectors or obstruct anomaly behaviors. Hybrid solutions propose a combination of techniques from detection and defense to protect web applications against XSS attacks at runtime. These solutions, generally, enable the detection of new attacks and act differently according to each observed behavior. Finally, attack prevention studies are focusing on the elimination of the source causes of the attack. This is mainly achieved through handling XSS vulnerabilities and can be divided into: \textit{mitigation}, \textit{detection}, \textit{protection} and \textit{hybrid}. Mitigation techniques are a set of guidelines or practices that should be preformed at early stages of the web applications development process. Vulnerability detection proposals enable checking the presence of insecure artifacts, in already developed web applications, prior deployment. Vulnerability protection techniques propose runtime actions to be taken in order to protect sensitive artifacts of web applications from being affected by malicious input data. Finally, vulnerability hybrid solutions enable dynamic detection and repair of applications vulnerabilities when they occur.  Figure~\ref{fig:categories} illustrates the distribution of studies regarding the set of distinguished categories. 

As shown in Figure~\ref{fig:categories}, the search process revealed much interest to XSS attack prevention compared with defense and detection. This is a positive fact since prevention is always better than cure. However, much attention is given to vulnerability detection instead of protection or hybrid. The less interest to XSS vulnerability mitigation techniques is understandable since those tasks require to seduce developers to respect security standards and follow strict security coding practices. Actually, this cannot easily be achieved due to time and cost constraints. The interest to vulnerability detection is also important since it is a first step toward attaining effective solutions. However, proposed solutions should be practical enough to be adopted by less experimented developers and respect time to market. This will be checked when answering RQ3.    
In the other side, XSS attack defense and detection are earning much attention compared with analysis and hybrid solution proposals. Runtime protection is important but more analysis studies are required to: (1) understand the peculiarity of new XSS attacks, (2) keep track of the development of hackers' tactics and (3) check the suitability of existing proposals to different types of XSS attacks. Table~\ref{tab:analysis} summarizes the focus of each analysis study and gives insights on their major findings.

	\begin{longtable}{lp{3.2cm}p{6.5cm}}
		\caption{Summary of analysis studies}
		\label{tab:analysis}\\      
		\hline\noalign{\smallskip}
		study & focus & summary of findings\\
		\noalign{\smallskip}\hline\noalign{\smallskip}
		Weinberger et al.~\cite{Weinberger2011} & evaluation of sanitization techniques adopted by 12 popular frameworks and web applications to defend against XSS attacks. &
		\begin{minipage} [t] {.4\textwidth} 
			\begin{itemize}
				\item limited support for context-sensitive auto-sanitization.
				\item no support for DOM-based XSS attacks that are dynamically generated at the client-side. 
				\item no appropriate sanitization to all available contexts where users should manually develop their own sanitizers.
			\end{itemize} 
		\end{minipage}\\
		
		\noalign{\smallskip}\hline\noalign{\smallskip}
		Scholte et al.~\cite{Scholte2012a} &  study the relationships between programming languages and vulnerabilities of web applications to XSS and SQL injections. &
		\begin{minipage} [t] {.4\textwidth} 
			\begin{itemize}
				\item PHP web applications are the most popular and the most vulnerable to XSS and SQL injections.
				\item SQL and XSS injections can often be prevented by enforcing common data types on input parameter types.
			\end{itemize} 
		\end{minipage}\\
		
		\noalign{\smallskip}\hline\noalign{\smallskip}
		Faghani and Nguyen~\cite{Faghani2013} & experimentation of analytical models that characterize the propagation of XSS worms in Online Social Networks (OSN). &
		\begin{minipage} [t] {.4\textwidth} 
			\begin{itemize}
				\item visiting friends more than strangers slows down the propagation. 
				\item the presence of large number of cliques in the network decreases the propagation speed.
				\item monitoring highly clustered structures of an OSN enables the containment of XSS worms within a community. 
				\item early detection of XSS worms can be achieved through monitoring a small number of users selected based on \textit{Degree} and \textit{PageRank} metrics. 
			\end{itemize} 
		\end{minipage}\\
		
		\noalign{\smallskip}\hline\noalign{\smallskip}
		Avancini and Ceccato~\cite{Avancini2013} &  comparison of genetic algorithms (GA) and concrete symbolic execution (CSE) as two test-case generation for the detection of XSS vulnerabilities. &
		\begin{minipage} [t] {.4\textwidth} 
			\begin{itemize}
				\item GA based test-case generation takes less time to generate a test case and parameter tuning has no significant impact. 
				\item CSE has higher coverage rate compared with GA due to the embedded sanity check. 
				\item an alternate composition of GA and CSE enables a trade-off between coverage rate and generation time. 
			\end{itemize} 
		\end{minipage}\\
		
		\noalign{\smallskip}\hline\noalign{\smallskip}
		Heiderich et al.~\cite{Heiderich2013} & study the prevalence of mutation XSS attacks (MXSS) and the suitability of existing techniques to defend against them. &
		\begin{minipage} [t] {.4\textwidth} 
			\begin{itemize}
				\item web-mailers are preferable targets to MXSS attacks.
				\item most existing defense techniques are far away from the effective detection against MXSS attacks.
			\end{itemize} 
		\end{minipage}\\
		
		\noalign{\smallskip}\hline\noalign{\smallskip}
		Bozic et al.~\cite{Bozic2015c} &  comparing the efficiency of IPOG and IPOG-F algorithms of combinatorial testing regarding the detection of XSS vulnerabilities. &
		\begin{minipage} [t] {.4\textwidth} 
			\begin{itemize}
				\item IPOG-F performs better than IPOG since it generates more comprehensive and sophisticated attack vectors. 
				\item adding constraints to the input parameter model significantly improves the quality of generated attack vectors by the two algorithms.
			\end{itemize} 
		\end{minipage}\\
		
		\noalign{\smallskip}\hline\noalign{\smallskip}
		Stock et al.~\cite{Stock2015} &  identify the root causes of client-side XSS attacks and classify them according to the complexity of their tainted-flows. &
		\begin{minipage} [t] {.4\textwidth} 
			\begin{itemize}
				\item 64\% of existing flaws are simple and only caused by the lack of security awareness of developers.
				\item 21\% quite complex flaws are caused by third-party libraries.
				\item 15\% complex flaws are caused by a combination of both causes.
			\end{itemize} 
		\end{minipage}\\

		\noalign{\smallskip}\hline\noalign{\smallskip}
		Zhang et al.~\cite{Zhang2015} & introduce XSS attacks caused by the use of insecure API implementations (XAS).&
		\begin{minipage} [t] {.4\textwidth} 
			\begin{itemize}
				\item the examination of 143 third-party applications for social networks showed that all are vulnerable to XAS attacks.
				\item a set of countermeasures may prevent XAS attacks. These include the need to set proper content-type header in API responses and sanitize user-input data incorporated in API responses.
			\end{itemize} 
		\end{minipage}\\
		
		\noalign{\smallskip}\hline\noalign{\smallskip}
		Weichselbaum et al.~\cite{Weichselbaum2016} & study the efficiency of Content-Security Policies (CSPs) in protecting web applications against XSS attacks. &
		\begin{minipage} [t] {.4\textwidth} 
			\begin{itemize}
				\item 94.72\% of real-world adopted CSPs can be bypassed. 
				\item relying on domain whitelists in CSPs is not sufficient to protect against trivial XSS attacks.
			\end{itemize} 
		\end{minipage}\\
		
		\noalign{\smallskip}\hline\noalign{\smallskip}
		Lin and Barcelo~\cite{Lin2016} & study the decidability of a logic conceived for analyzing Mutation XSS (MXSS). &
		\begin{minipage} [t] {.4\textwidth} 
			\begin{itemize}
				\item regular expressions alone are not sufficient for the detection of security vulnerabilities. 
				\item a sound logic is required for expressing constraints useful for the analysis of MXSS attacks and vulnerabilities.
			\end{itemize} 
		\end{minipage}\\
		
		\noalign{\smallskip}\hline\noalign{\smallskip}
		Bazzoli et al.~\cite{Bazzoli2016} &  derive a set of recommendations for tackling the limitations of blackbox vulnerability scanners in the detection of reflected XSS vulnerabilities. &
		\begin{minipage} [t] {.4\textwidth} 
			\begin{itemize}
				\item use a diverse set of distinct payloads instead of fuzzing or mutating a small set of arbitrary selected samples. 
				\item adopt a context-sensitive based selection of candidate payloads.
				\item select short payloads with a small character set. 
				\item regularly update the set of payloads for the detection of new vulnerabilities.
			\end{itemize} 
		\end{minipage}\\

		\noalign{\smallskip}\hline\noalign{\smallskip}
		Lekies et al.~\cite{Lekies2017} & examine the ability of existing defense techniques to defend against XSS attacks caused by leveraging script gadgets (CR-XSS). &
		\begin{minipage} [t] {.4\textwidth} 
			\begin{itemize}
				\item the examination of 10 XSS mitigation tools (HTML sanitizers, filters, web application firewalls and Content-Security policies) showed that they are unable to handle CR-XSS attacks.
			\end{itemize} 
		\end{minipage}\\
		
		\noalign{\smallskip}\hline\noalign{\smallskip}
		Melicher et al.~\cite{Melicher2018} &  check the prevalence and identify the causes of DOM XSS vulnerabilities. &
		\begin{minipage} [t] {.4\textwidth} 
			\begin{itemize}
				\item 83\% of vulnerabilities come from advertising and analytics domains. 
				\item DOM XSS vulnerabilities can be eliminated by blocking Ads. 
				\item DOM XSS vulnerabilities are also caused by the use of incorrectly implemented templates from templating frameworks.
			\end{itemize} 
		\end{minipage}\\
		
		\noalign{\smallskip}\hline\noalign{\smallskip}
		Steffens et al.~\cite{Steffens2019} &  check the prevalence of persistent client-side XSS vulnerabilities. &
		\begin{minipage} [t] {.4\textwidth} 
			\begin{itemize}
				\item more than 8\% of the examined 5,000 highest-ranked sites are found vulnerable to persistent client-side XSS attacks.
				\item 4 distinct scenarios are identified showing how client-side storage is used in an insecure manner.
			\end{itemize} 
		\end{minipage}\\		
		
		\noalign{\smallskip}\hline\noalign{\smallskip}
		Wijayarathna and Arachchilage~\cite{Wijayarathna2019} & check the usability of OWASP ESAPI sanitizers to prevent XSS attacks from developers' perspectives. &
		\begin{minipage} [t] {.4\textwidth} 
			\begin{itemize}
				\item programmers are unable to identify all the locations where sanitizers should be instrumented. They also accidentally encode inputs instead of outputs. The problem is due to the absence of detailed documentations with enough comprehensive examples.
				\item programmers use wrong encoding methods to encode data due to the absence of validation routines.
			\end{itemize} 
		\end{minipage}\\		
		
		\noalign{\smallskip}\hline\noalign{\smallskip}
		Chaliasos et al.~\cite{Chaliasos2019} &  experiment a variant of XSS attacks that are caused by slight modifications on the AST of web pages (JSM-XSS).  &
		\begin{minipage} [t] {.4\textwidth} 
			\begin{itemize}
				\item JSM-XSS can easily bypass whitelist scripts based defense mechanisms.  
			\end{itemize} 
		\end{minipage}\\
		
		\noalign{\smallskip}\hline\noalign{\smallskip}
		Schuckert et al.~\cite{Schuckert2020} & extract code patterns that harden static analyzers' detection of XSS vulnerabilities. &
		\begin{minipage} [t] {.4\textwidth} 
			\begin{itemize}
				\item 19 source code patterns are extracted from open source projects and CVE reports. These include the use of super global variables such as \texttt{\$\_SERVER} and \texttt{foreach} on super global variables.
				\item there may still be undetectable code patterns.
			\end{itemize} 
		\end{minipage}\\		
		
		\noalign{\smallskip}\hline\noalign{\smallskip}
		Buyukkayhan et al.~\cite{Buyukkayhan2020} & capture the development of reflected XSS exploits over 10 years of time.  &
		\begin{minipage} [t] {.4\textwidth} 
			\begin{itemize}
				\item most reflected server XSS exploits are clear and obfuscation is rarely used. 
				\item the majority of exploits could be blocked by existing filtering based defense techniques available at browsers before reaching the server. 
				\item few complex exploits are hard to be detected by existing defense systems. 
			\end{itemize} 
		\end{minipage}\\
		
		\noalign{\smallskip}\hline\noalign{\smallskip}
		Bui et al.~\cite{Bui2020} & study XSS vulnerabilities in cloud-application add-ons. &
		\begin{minipage} [t] {.4\textwidth} 
			\begin{itemize}
				\item 9\% of analyzed add-ons are found vulnerable to XSS.
				\item a set of countermeasures are proposed to tackle the issue. These include: implement the add-on logic on the server instead of the client-side and prevent sharing access tokens to delegate all user permissions.
			\end{itemize} 
		\end{minipage}\\		
		
		\noalign{\smallskip}\hline\noalign{\smallskip}
		Talib and Doh~\cite{Talib2021} & evaluate and compare the performance of 12 dynamic open-source XSS filters.&
		\begin{minipage} [t] {.4\textwidth} 
			\begin{itemize}
				\item filters are not sufficient alone for the detection of XSS attacks.
				\item tested filters work well with malicious scripts but not with benign ones (generate high false positives).  
				\item not suitable for the detection of DOM-based XSS (DXSS). 
				\item no guarantee that filters work properly with other contexts than HTML. 
			\end{itemize} 
		\end{minipage}\\
		
		\noalign{\smallskip}\hline
	\end{longtable}

\section{What type of XSS attacks are addressed by contemporary studies? (RQ3)}
\label{sec:rq3}
\begin{figure}[H]
	\centering
	\includegraphics[width=.9\textwidth]{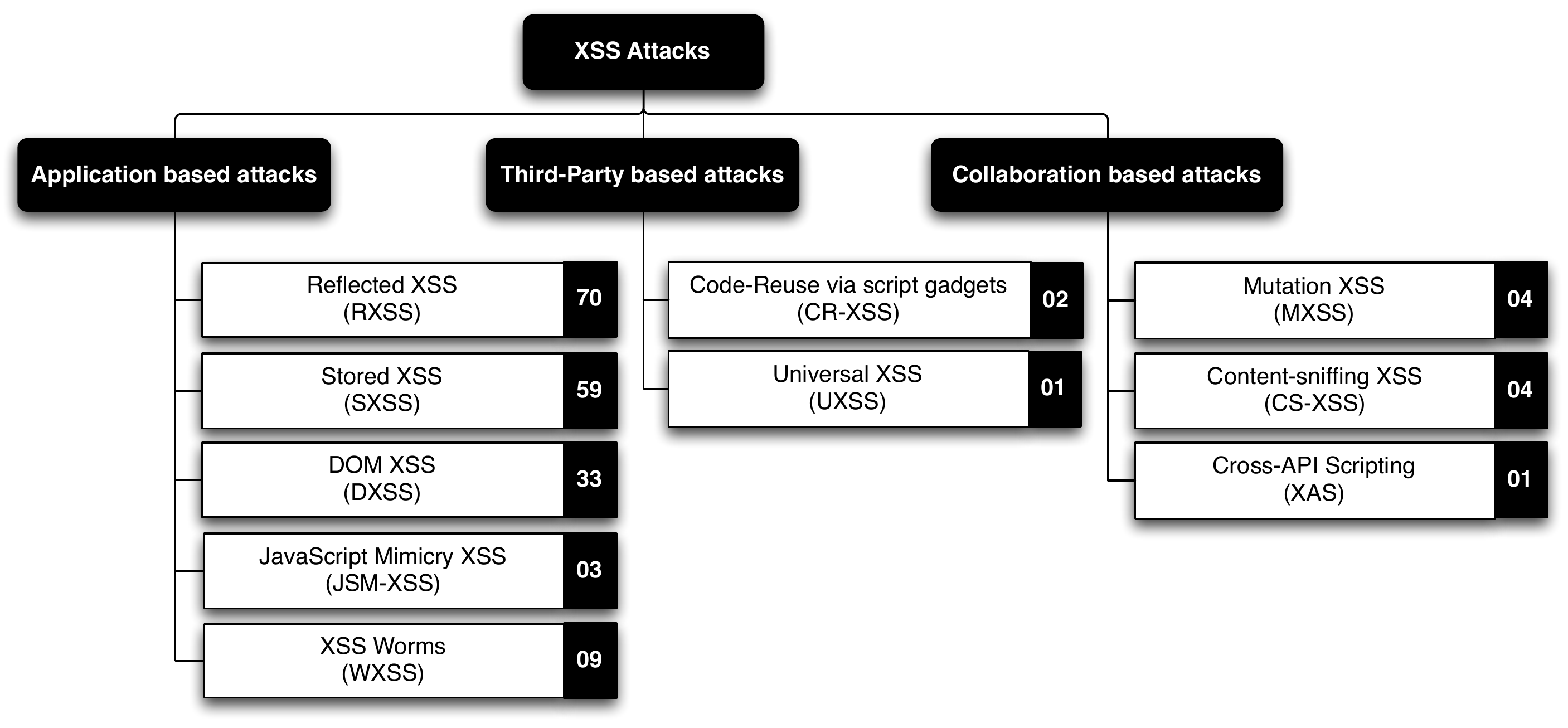}
	\caption{Types of XSS attacks}
	\label{fig:attackstax}
\end{figure}

Besides the three basic types reported by OWASP~\cite{OWASP}, other variants of XSS attacks are also covered and addressed by contemporary studies. A common adopted classification is the one reported by OWASP itself that considers the location where untrusted data are supplied and processed: \textit{client-side} and \textit{server-side} XSS. This classification is adopted by several included studies~\cite{Stock2015, Yamazaki2018, Steffens2019, Musch2019, Bui2020, Tariq2021}. In this review, we propose a more comprehensive classification based on the source of the vulnerability causing the attacks. Accordingly, three categories can be distinguished: \textit{Application-based}, \textit{Third-party-based} and \textit{Collaboration-based}. Figure~\ref{fig:attackstax} shows the proposed classification of XSS attacks together with the number of studies addressing each attack. In the sequel we give a detailed description of each category with the list of their included attacks.

\subsection{Application-based XSS attacks} Attacks of this category are mainly caused by web applications themselves. Inexperienced developers with security issues often omit the integration of proper sanitization to every user controlled sources. Five of existing attacks fall in this category including the three basic OWASP attacks reported in~\cite{OWASP}:

\begin{enumerate}
	\item \textit{Reflected XSS (RXSS)}: RXSS attacks occur when malicious data supplied as part of HTTP requests become part of server responses without being properly sanitized. Consequently, malicious data embedded in HTTP responses reach back the client-side and get executed when rendered by browsers. Typical targets of RXSS are web applications with search capabilities, where embedded malicious data become part of search results or error messages. These attacks are also referred to as non-persistent or Type I~\cite{OWASP}. They are said to be non-persistent because malicious data are not permanently stored in the server.
	
	\item \textit{Stored XSS (SXSS)}: SXSS attacks appear when malicious data are supplied in input forms to be stored in server application databases or files without being sanitized. This way, users requesting data from those applications are supplied with server responses containing malicious data that become harmful when being rendered by browsers. Typical targets of SXSS attacks are online social web applications and forums where malicious data can be posted, stored in databases and hence infect every user accessing them. SXSS attacks are also referred to as persistent or Type II~\cite{OWASP}. They are persistent because malicious data stored in the server remain harmful and effective until being detected, removed or filtered.
	
	\item \textit{DOM XSS (DXSS)}: DXSS attacks are render-time attacks. Contrary to RXSS and SXSS, malicious data used in DXSS attacks are used to dynamically alter the DOM trees generated by browsers at rendering phase. Attackers use DXSS attacks for web applications that accept data from user-controlled sources to be used as specific DOM objects values. The peculiarity of DXSS is that malicious data can be embedded in URLs as values to specific DOM objects or HTML elements but never reach the server. These attacks are also referred to as Type 0 and are firstly reported in 2005 by Amit Klein~\cite{Klein2005}.
	
	\item \textit{JavaScript Mimicry XSS (JSM-XSS)}: Instead of injecting malicious scripts, attackers make use of scripts already used in web applications to launch JSM-XSS attacks. It has been reported and demonstrated with several examples in~\cite{Wagner2002} that legitimate scripts injected in different places than those intended by developers may lead to harmful behaviors of web applications. JSM-XSS attacks are hard to be detected and easily pass whitelisting based filters. 
	
	\item \textit{XSS worms (WXSS):} WXSS attacks are XSS attacks with self-replication capability. RXSS and SXSS attacks with self-replication capability are classified as WXSS. This distinction is adopted since specific other type of web application vulnerabilities are required for spreading attacks and become worms. WXSS attacks are more dangerous since they propagate among web application users and progressively infect other users over time and remain so until being detected and removed. Typical targets of such attacks are online social network applications where user profiles are somehow linked to each others. To launch a WXSS attack, web application vulnerabilities are used to infect the first user and inject a malicious script; the execution of the injected script uses other application vulnerabilities to escalate the privilege and perform some actions on the user’s behalf. Those actions enable self-replication and hence cause the infection of other linked users.
\end{enumerate}

\subsection{Third-party based XSS attacks} The attacks of this category do not directly rely on web application vulnerabilities, instead, they rely on third-party vulnerabilities such as browsers, browser extensions and third-party libraries or frameworks used by web applications. Two  attacks found reported in the literature fall in this category:

\begin{enumerate}
	\item \textit{Code-Reuse via Script Gadget (CR-XSS)}: Triggering a CR-XSS attack basically requires a knowledge of the scripts included in the libraries and/or frameworks used by targeted applications. Those are also known as script gadgets. A successful CR-XSS attack is achieved by injecting HTML codes with disguised payloads (i.e., in non-executable form) that match DOM elements and provoke the execution of script gadgets. Apparently benign payloads are transformed into executable scripts by the execution of a single or chain of gadgets. CR-XSS is introduced in 2013 by Lekies et al.~\cite{Lekies2017}. Script gadgets can also be sourced from user-land code, in this case, CR-XSS attacks become web application based-attacks. 
	
	\item \textit{Universal XSS (UXSS)}: UXSS is caused by the lack of proper sanitizations of URLs by the browser itself or one of its extensions. For triggering a UXSS attack, intruders get benefits of vulnerabilities located in browsers or browser extensions. Therefore, by seducing users to click in a link that provokes the execution of installed plugins in the visitor browser, the plugin triggers the execution of malicious scripts incorporated in the links. They are called universal since malicious scripts can be executed in the context of any web site and not directly related to specific web applications.
	
\end{enumerate}

\subsection{Collaboration based XSS attacks} Attacks of this category can only be succeeded with the presence of web applications and third-party vulnerabilities. If any of those vulnerabilities is missing, those attacks fail. Three of the reported attacks in the literature fall in this category:

\begin{enumerate}
	\item \textit{Mutation XSS (MXSS)}: MXSS attacks are mainly caused by the capability of browsers to transform formatted HTML strings into valid DOM elements making use of the \texttt{innerHTML} property used in web applications. Attackers supply their malicious data as formatted HTML strings associated as values to the \texttt{innerHTML} property in vulnerable target web applications. Those malicious data may bypass application filters and transformed by browsers into valid HTML contents, they are inserted as new DOM elements and then executed as part of rendered pages. Typical targets of MXSS are webmail applications with HTML content message transmission capabilities. Transmitted messages are transformed (i.e., mutated) into valid scripts executed by the receiving browsers at rendering the content of messages. MXSS attacks are firstly reported in 2017 by Heiderich et al.~\cite{Heiderich2013}.
	
	\item \textit{Cross-API Scripting (XAS)}: XAS is a kind of XSS attacks that targets web applications providing Restful APIs to third-party developers such as social networks. Attackers inject malicious scripts in their own application profiles and hence users of third-party applications become vulnerable to XAS. Malicious data are transmitted to victims' browsers through using APIs to retrieve data from web applications. The root cause of XAS attacks is the lack of proper sanitization of data by: (1) the web application itself that permits attackers to inject malicious data, (2) third-party web applications for their acceptance of malicious API responses without proper sanitization. This causes the execution of malicious data at the victims' browsers. The term XAS is firstly introduced in 2013 and later on 2015 by Zhang et al.~\cite{Zhang2015}.
	
	\item \textit{Content-Sniffing XSS (CS-XSS)}: XSS attacks caused by the misinterpretation of file content types by browsers are named content-sniffing XSS and referred to in this paper as CS-XSS. To conduct a CS-XSS attack, malicious data are supplied as part of separate media files (e.g., PDFs, images) uploaded to vulnerable web applications. Those files become harmful to every user loading them in a vulnerable browser. To be successful, CS-XSS attacks require web applications permitting the upload of infected files and browsers using a content-sniffing practice to deduce the type of files. With the presence of scripts in files, browsers consider them as HTML files and hence get rendered where the injected scripts are executed at the victims' browsers.
\end{enumerate}

\begin{figure}[H]
	\centering
	\includegraphics[width=.5\textwidth]{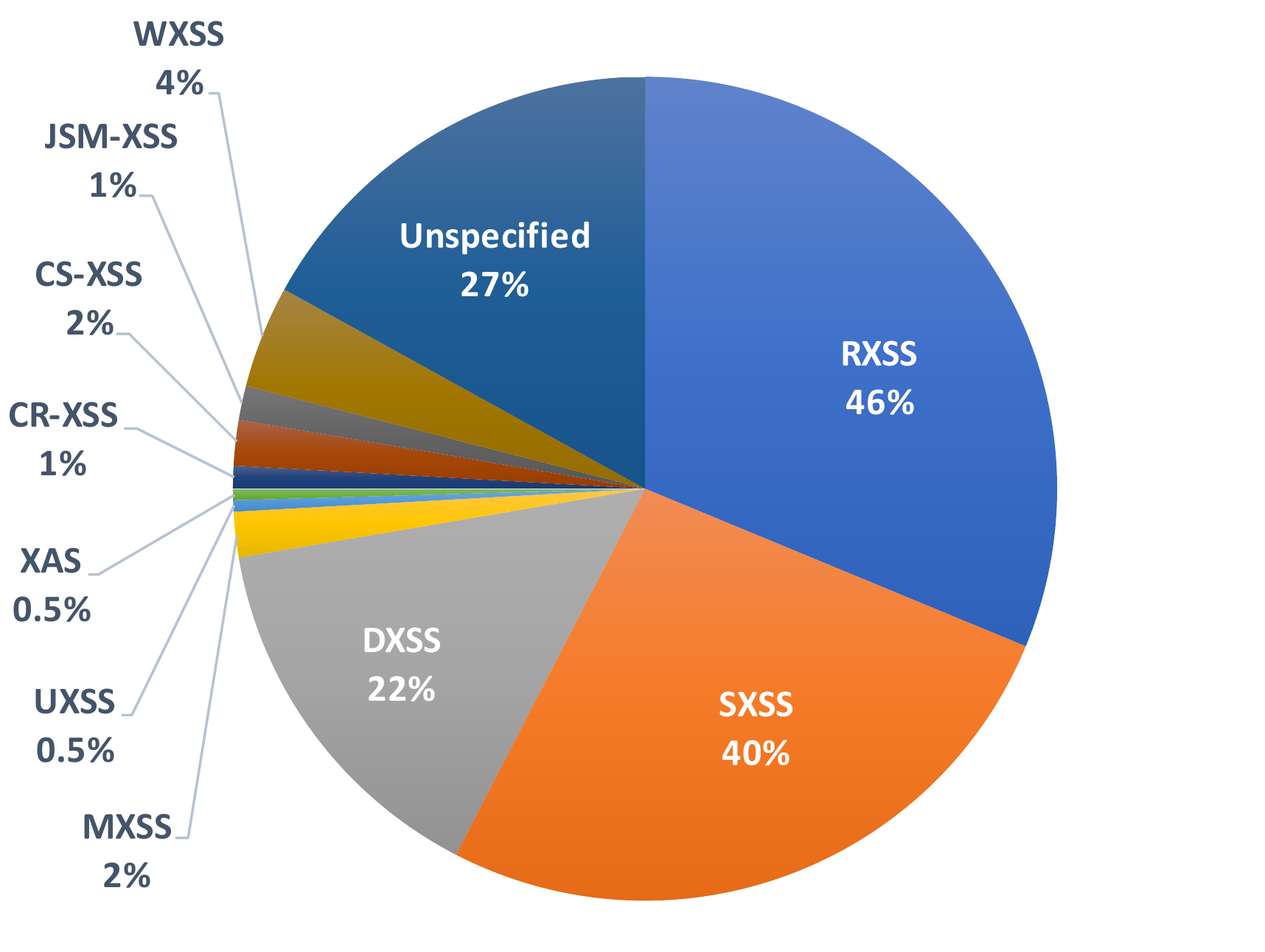}
	\caption{Distribution of included studies per addressed XSS attack type}
	\label{fig:attackscore}
\end{figure}

Figure~\ref{fig:attackscore} shows the distribution of included studies regarding the different XSS attack types. Application-based XSS attacks are the most addressed, they are reported and handled by 98 studies individually, together or with other attack types. Specifically, RXSS is the most referred attack followed by SXSS and DXSS. The focus on RXSS is understandable since it is the most common and the easiest to be detected compared with other attacks. Remarkably, DXSS is getting much attention compared with the results reported in 2015 by Hydara et al.~\cite{Hydara2015}. The other attacks are less addressed. Specifically, XAS and UXSS are only addressed once compared with other variants. The problem is due to the fact that those attacks target specific applications or platforms. XAS attacks target online social network applications with Restful APIs and UXSS targets specific browsers and browser plugins. Through the data extraction process, we found that studies with unspecified attacks are mostly referring to basic attacks, specifically RXSS and/or SXSS, however to be concise, we separated them and we did not consider them as basic attacks. 

\begin{figure}[H]
	\centering
	\includegraphics[width=1\textwidth]{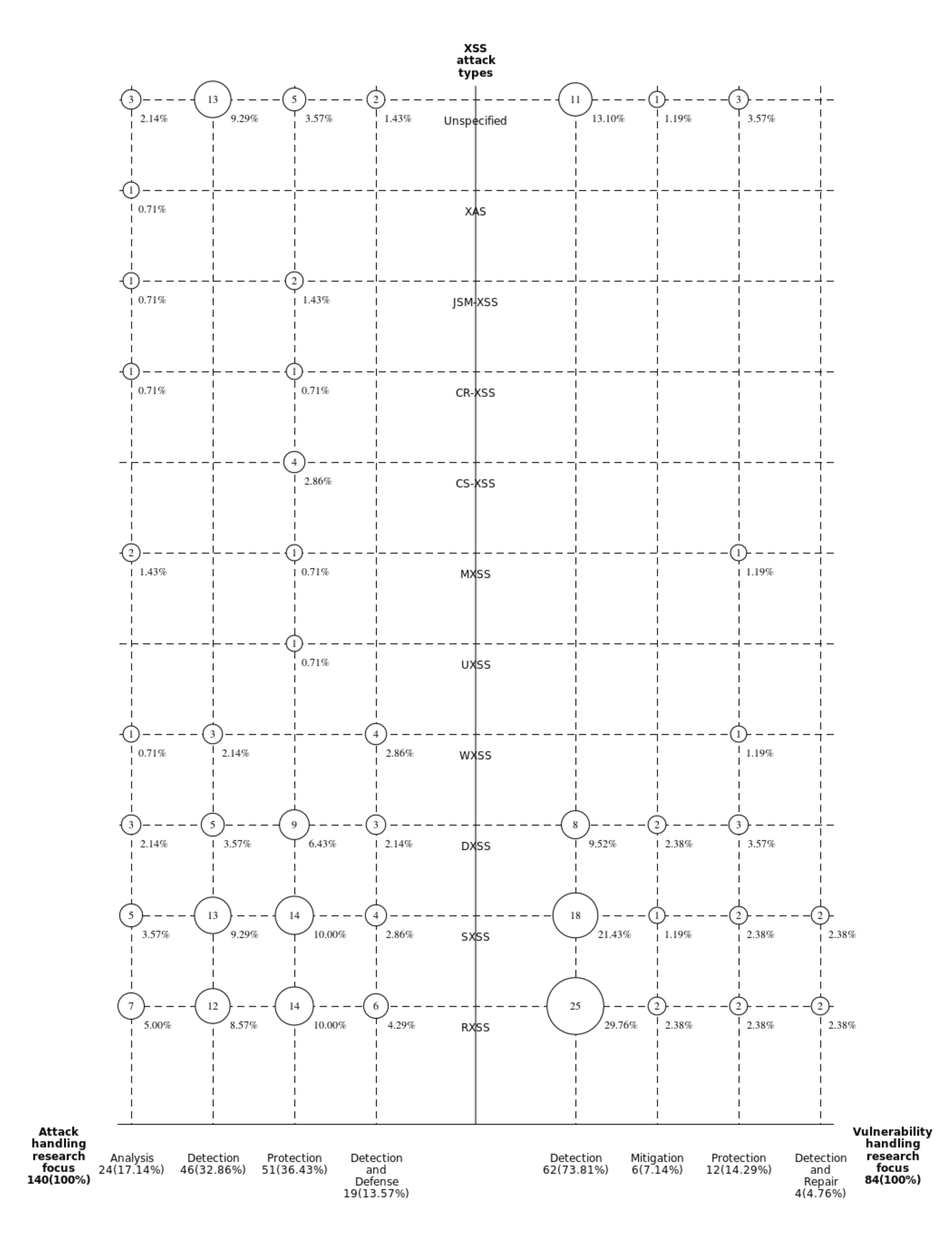}
	\caption{Distribution of the research focus of included studies regarding the different type of XSS attacks}
	\label{fig:map1}
\end{figure}

Figure~\ref{fig:map1} maps the type of included studies to XSS attack types. The figure clearly shows a bias toward basic attacks (i.e., RXSS, SXSS and DXSS). Other XSS attack variants are rarely studied. Specifically there is a noticeable research gap regarding the prevention of such attacks.

\section{What techniques have been used to tackle the XSS issue? (RQ4)}
\label{sec:techniques}
\begin{figure}[H]
	\centering
	\includegraphics[width=1\textwidth]{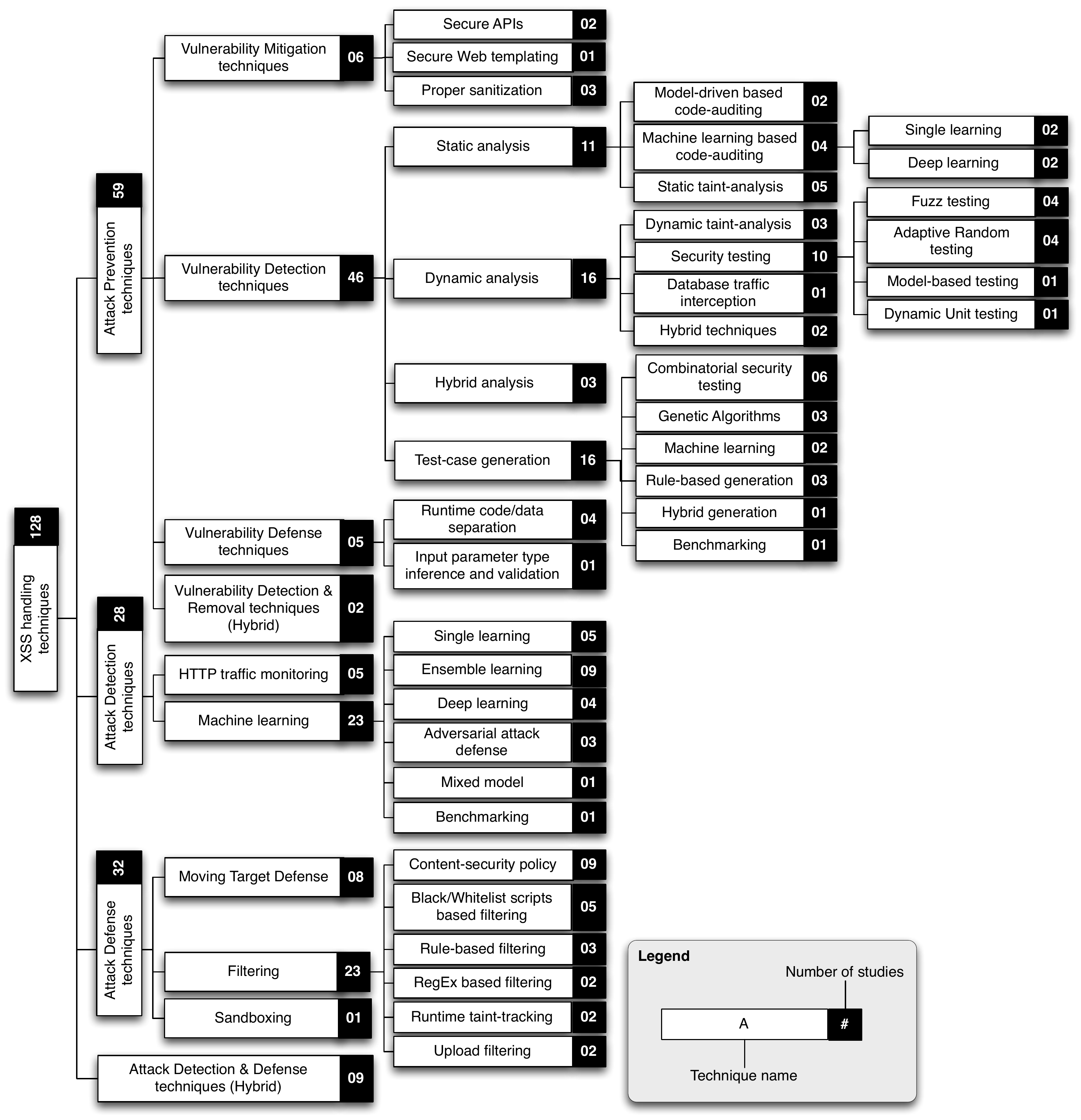}
	\caption{Taxonomy of proposed solutions to XSS attacks}
	\label{fig:solutions}
\end{figure}

Different techniques have been proposed for handling XSS attacks. Those techniques can be categorized following their intents and natures. Figure~\ref{fig:solutions} gives an overview of the proposed classification of solution proposals together with the number of included papers addressing each technique. 

\subsection{XSS Prevention techniques}
\label{sec:prevention}
Prevention techniques refer to the set of security measures that prohibit the appearance of attacks in the first place and enable the development of XSS-free web applications. Those measures need to be taken at early stages of the development process for Web applications, specifically, at the construction phase and prior deployment. They can be divided into XSS vulnerability mitigation, detection and/or defense techniques.

\subsubsection{XSS vulnerability mitigation techniques} XSS attacks are mainly caused due to the absence of proper sanitization of user-controlled inputs to web applications. In order to reduce the chances of being attacked, several measures can be taken by web application developers.
The study of the literature shows three main secure coding practices: use of \textit{secure API}s, use of \textit{secure web templating} and instrumentation of \textit{proper sanitizer routines}. In the sequel, we provide a brief description of each proposal and discuss its limitations.

\paragraph{\textbf{Use of secure APIs:}} Browsers often come with built-in APIs that enable performing complex operations, such as manipulating the DOM structure of web documents, in a high-level manner. Those APIs are not without flaws; they can be used by intruders to launch DXSS attacks.   
Wang et al.~\cite{Wang2021} proposed developing secure web applications by forbidding direct calls to vulnerable JavaScript and TypeScript engines. Instead, secure APIs need to be developed, integrated and made available for web developers. The new APIs enforce typing of HTML element values to prevent any future script injection. However, new APIs also inherit the unsoundness of JavaScript compilers. In addition, learning new APIs requires much efforts from developers which is a non cost-effective solution for large development organizations and wrokflows.
Musch et al.~\cite{Musch2019} proposed a lightweight solution that consists of using DOM API wrappers to prevent the injection of scripts through calling built-in DOM API functions. Occasionally, if values to DOM APIs are required from third parties, those values are redirected to an API wrapper for sanitization before being forwarded to standard APIs. Unfortunately, the proposed scheme is unable to wrap all JavaScript sinks such as \textit{eval()} and has a compatibility issue to be automatically integrated to all web applications. Consequently, users need to be interfered to manually add wrapper scripts at the top of each web page which is a tedious task. 

\paragraph{\textbf{Use of secure web templating:}} The use of Web templating is a common practice that enables web designers and developers to generate their customized web applications through reconfigurations of already designed templates. 
Samuel et al.~\cite{Samuel2011} proposed the construction of secure web applications making use of web templating frameworks with embedded auto-sanitization mechanisms. The proposed system parses templates, infers input variable types and annotates them with context type qualifiers. Annotated templates are then compiled into the target language of pages with instrumented sanitizers from a sanitizer library. The solution is experimented only with Google Closure framework templates and unable to determine all variable contexts due to browser quirks.

\paragraph{\textbf{Use of proper sanitization routines:}} Instrumenting sanitizers at all sensitive injection points is an intuitive solution to prevent XSS attacks. 
Huang et al.~\cite{Huang2004} proposed a sanitization tool named WebSSARI. The proposed tool can be used by web developers to perform automatic static taint-tracking; tainted data reaching sensitive sinks are treated by automatic integration of sanitizers from a database. The reliance to a predefined set of sanitizers is the main drawback of the approach and the experiments conducted by the authors showed that the tool produces large number of false positives (26.9\%). 
Heiderich et al.~\cite{Heiderich2017} proposed a ready to use sanitizer named DOMPurify. It enables transforming malicious HTML strings into safer versions by striping out every dangerous and confusing characters and thereby prevent XSS attacks. It also prevents bypassing through encryption where scripts are firstly decrypted and filtered prior rendering. However, DOMPurify can easily be passed by CR-XSS attack vectors. In addition, DOMPurify need to instrumented manually at each injection point which is not feasible in practice.
To enrich user-libraries with correct and effective sanitizers, Hooimeijer et al.~\cite{Hooimeijer2011} proposed a domain specific language (BEK) to describe the behavior of sanitizers. BEK descriptions can be transformed into a special finite state machines to be formally checked for their correctness and effectiveness against XSS attacks. BEK can detect some formal properties of single and composed sanitizers such as commutativity and idempotence. Correct sanitizers can automatically be transformed into Java or C\# for being integrated into existing systems. Unfortunately, tested santizers need first to be transformed manually to BEK which is a tedious and error-prone task.

\subsubsection{XSS vulnerability detection techniques} Missing flaws by secure-coding practices can be detected using appropriate approaches and tools for scanning web applications prior deployment. The literature is rich with vulnerability detection techniques that can be classified into four main categories: \textit{static}, \textit{dynamic}, \textit{hybrid analysis} and \textit{test-case generation}. 

\paragraph{\textbf{Static analysis}} Static analysis approaches and tools focus on exploring web application source codes, without their execution, to find security vulnerabilities. They have the ability to discover all the execution paths which effectively reduces the false negative rate. Three main static analysis approaches are proposed by included studies: \textit{model-driven based}, \textit{machine-learning based} code auditing and \textit{static taint-analysis}:

\begin{itemize}
	\item \textbf{Model-driven based code auditing:} web applications vulnerabilities can statically be checked through the analysis of their derived abstract models. This technique is explored by two included studies~\cite{Shar2012, Vernotte2014}. In~\cite{Shar2012}, the defense measures instrumented by developers are recovered and checked for correctness and consistency. For this sake, web application pages are transformed into control flow graphs. Nodes of generated graphs are classified based on security measures instrumented by developers on each node. A set of well-defined rules are checked to verify the adequacy of used defense measures in preventing XSS attacks on those nodes. The solution requires manual specifications of escaping APIs used in web applications; this makes it unsuitable for large web applications.
	In~\cite{Vernotte2014}, the behavior of web applications is modeled using a well-defined domain specific language (DSL) making use of functional specifications of web applications. Test-patterns are explicitly described using natural language and automatically transformed into formalized forms named test-purposes. Test-cases are then generated from test-purposes and executed on the abstract behavioral model for verification. This solution is also not suitable for large applications since it requires user interventions for the description of test models using the proposed DSL.
	
	\item \textbf{Machine learning based code auditing:} taking benefits from already discovered and known vulnerabilities, machine learning can be used as a code auditing approach to check the vulnerability of developed web pages against XSS attacks. The approach is found being explored by four included studies~\cite{Shar2013, Gupta2016c, Li2020a, Maurel2021}. Datasets constituted of safe and vulnerable web pages are collected from different sources such as GitHub\footnote{GitHub home page: \url{https://github.com}} and Common Weakness Enumeration (CWE)\footnote{CWE: Common Weakness Enumeration: \url{https://cwe.mitre.org}}. Features are extracted from web page sources through feature engineering like in~\cite{Shar2013, Gupta2016c, Li2020a} or through feature learning like in~\cite{Maurel2021}. Feature vectors are built and fed to 
	machine learning classifiers for training and then for classifying web pages into vulnerable and safe. Two types of machine learning are found used in the literature: single learning in~\cite{Shar2013, Gupta2016c} and deep learning in~\cite{Li2020a, Maurel2021}. Specifically, a variant of decision tree classifier (J48) is used in~\cite{Gupta2016}, Multi-layer Perceptron classifier (MLP) is used in~\cite{Shar2013, Maurel2021} and Bidirectional Long short-term memory network (BLSTM) is used in~\cite{Li2020a}. Those solutions are language dependent (PHP) and produce high false positive rates.
	
	\item \textbf{Static taint-analysis:} consists of tracking input data from sources to sensitive sinks in web application source codes without being executed. An XSS vulnerability is reported if an input data reaches a sensitive sink without being properly sanitized. Jovanovic et al.~\cite{Jovanovic2006a} developed a typical tool named Pixy that enables static taint-analysis through exploring automatically generated control-flow graphs. Pixy is designed for the detection of RXSS vulnerabilities in PHP web applications. Wang et al.~\cite{Wang2011} extended Pixy for the support of SXSS vulnerabilities. The proposed algorithm in~\cite{Wang2011} uses both data-dependence and control-dependence graphs to identify program slices causing SXSS vulnerabilities. However, Pixy is a language dependent tool and does not support PHP object-oriented features. In addition, it produces high false positive rates (around 50\%) and is unable to detect flaws across multiple pages. Steinhauser and Gauthier~\cite{Steinhauser2016} developed JSPChecker which is an another tool implementing static taint-analysis for the detection of XSS vulnerabilities. Instead of tracking input data to sinks, JSPChecker tracks the list of sanitizer routines applied to input data before reaching sinks in order to discover context inconsistencies. The solution requires manual identifications of sanitizers and mapping them to safe output contexts. Wassermann and Su~\cite{Wassermann2008} used static taint-analysis to check the ability of vulnerable inputs sent by users to invoke the JavaScript interpreter. The solution is browser dependent and cannot handle arbitrarily complex and dynamic codes. Yan and Qiao~\cite{Yan2016} proposed an optimized algorithm for static taint-tracking through reverse code auditing. Instead of exploring all existing paths, only few and relevant paths are need to be explored for the detection of vulnerabilities. However, insufficient experiments are conducted to check the correctness of the proposed algorithm. 
	
\end{itemize} 

\paragraph{\textbf{Dynamic analysis}} Contrary to static analysis, dynamic analysis approaches and tools aim at discovering flaws of web applications while they are running. They enable testing web applications on real-time scenarios. Dynamic analysis methods focus on information acquired at runtime. They detect the presence of an XSS vulnerability based on HTTP responses by dynamically sending requests to servers. Dynamic analysis methods do not rely on the presence of web application codes and have generally lower false positive rates. Several dynamic techniques are used in the literature. In the sequel, we describe each of those approaches and we discuss the correspondent studies included in this review.

\begin{itemize}
	\item \textbf{Dynamic taint-analysis:} consists of marking input sensitive data and track their propagation when tested applications are running. A vulnerability is reported when input data reach predefined program locations or sinks. Martin and Lam~\cite{Martin2008} proposed a taint-analysis technique armed with a model checker to prove the success of attacks. The proposed system is designed to handle taint-based vulnerabilities including XSS and SQL injections. Tested vulnerabilities are described using a specific language named PQL~\cite{Martin2005}. Users should provide the PQL specifications of tested vulnerabilities together with a set of input query parameter values. The proposed system explores the space of possible input tests and monitors the execution of the application under each input. The model checker checks whether input tests successfully reach specified sinks and reports the correspondent attack paths. The efficiency of this solution relies on the correctness of specifications given by developers. It fails at the detection of vulnerabilities with the presence of advanced sanitizers and it is only applicable to Java based web applications. 
	Steinhauser and Tuma~\cite{Steinhauser2019} developed a tool that uses an extended dynamic taint-tracker for the detection of browser-context-sensitive XSS flaws. They are vulnerabilities caused by the use of inappropriate sanitizers for particular browser contexts. The proposed tool performs a dynamic taint-tracking to identify the set of sanitizers applied to tainted values. Afterwards, a model browser checks the resulted HTML responses to determine the browser contexts where tainted data are passed through. A set of well-defined rules are used to check the compatibility of used sanitizers with identified browser contexts. The proposal is not suitable for the detection of DOM-based vulnerabilities and is applicable only to Django based web applications.
	
	Instead of monitoring program execution, taint-inference matches input values from HTTP requests with those resulted from their correspondent HTTP responses and infers the taint-flow between them. Any match is reported as an XSS vulnerability. Pan et al.~\cite{Pan2016b} proposed a new taint-inference technique for the detection of web applications vulnerability to RXSS attacks. To overcome the effects of URL rewriting and HTML sanitization on matching results, the authors used local sequence alignment and removal gap penalty inspired from molecule sequence alignment used in bio-informatics~\cite{Gusfield1997}. The technique performs poorly when tags are removed by HTML sanitizers.  
	
	\item \textbf{Security testing:} Security testing refers to the set of techniques providing evidences that web applications are safe and reliable, and that they do not accept unauthorized inputs. They test the impact of malicious or unexpected inputs on their  functionalities when they are running. Several techniques have been used for the detection of XSS vulnerabilities:
	
	\begin{itemize}
		\item \textbf{Fuzz testing: } consists of discovering web application vulnerabilities by injection of invalid, malformed, or unexpected inputs and observation for exceptions on their behaviors. This technique is explored by four included studies~\cite{McAllister2008, Duchene2013, Duchene2014, Eriksson2021}. McAllister et al.~\cite{McAllister2008} proposed a guided fuzzer by a collected set of use-cases from the server. The fuzzer (w3af) replaces the input parameter values of real-world requests with malicious strings from a database. The solution is only experimented with Django web applications. Duchène et al.~\cite{Duchene2013} proposed the use of a guided w3af fuzzer with an inferred control and data-flow model to increase the capability of detecting XSS vulnerabilities. The solution does not support all encoding functions and requires resetting repeatedly tested applications to their initial states.
		In a recent work~\cite{Duchene2014}, the authors replaced the w3af fuzzer with more appropriate one that uses well-defined attack grammars to transform benign inputs into fuzzed inputs. The new solution requires the intervention of users to identify non-deterministic values of specific attributes related to the implemented tool.
		Eriksson et al.~\cite{Eriksson2021} proposed a black-box scanner that loads pages in a modified browser and observes their execution flows. Fuzzed inputs are sent and the scanner detects whether they have been executed. The solution fails to detect vulnerabilities when too much randomness is used on inputs.
		
		\item \textbf{Adaptive random testing:} instead of testing web applications using arbitrary set of test-cases, adaptive random testing can be adopted to select more distinctive and likely successful test-cases for each step~\cite{Huang2021}. Lv et al.~\cite{Lv2019} adopted a distance based selection of test-cases. 
		Rocha and Souto~\cite{Rocha2014} proposed to test web applications on generated test-cases according to contexts and qualifiers associated to each discovered vulnerable point. 
		Leithner et al.~\cite{Leithner2021} proposed a feedback based approach for the selection of next injected test-cases. Therefore, based on the current injected vector and the reached output context, the next attack vector is an update of the current vector that likely reaches a number of defined ideal locations making use of weights. Those weights are required to be selected by developers which may considerably affect the performance of the proposed system.
		Tripp et al.~\cite{Tripp2013} used reinforcement learning for the selection of likely successful attack vectors. Using a database containing 500 million payloads and through a learning capability from failed attacks and previously selected payloads, the algorithm selects, in each step, a payload that likely bypass instrumented sanitizers in web applications. The solution produces false negatives with the presence of sanitizers that limit input lengths or trim a fixed number of characters from the head or the tail of input strings.  
		
		\item \textbf{Model based testing:} in such kind of techniques, trained models are used to distinguish vulnerable from non vulnerable pages. Avancini and Ceccato~\cite{Avancini2013a} developed a test-oracle named Circe for testing XSS vulnerabilities of web applications. As a test oracle, a model is constructed based on the structure of web pages generated using safe inputs. Malicious inputs from a library are injected and resulted page structures are matched to those generated from safe inputs. Any deviation is reported as a potential vulnerability. The model is unable to detect all real-world scenarios due to the adopted coverage criterion and its reliance to a fixed set of malicious inputs.
		
		\item \textbf{Dynamic unit testing:} unit testing is another dynamic analysis technique used for vulnerability detection. In dynamic unit testing, a program unit is executed in isolation with selected inputs, the results are then compared with the expected outcomes. Mohammadi et al.~\cite{Mohammadi2017} proposed the use of a dynamic unit testing approach for the detection of XSS vulnerabilities in JSP web applications. Unit tests for each page are automatically constructed and executed using a unit test execution framework. For test inputs, a well-established attack grammar is used for the generation of proper inputs.
	\end{itemize}
	
	\item \textbf{Database traffic interception:} providing a control interface between web applications and their databases enables XSS analyzers to detect SXSS vulnerabilities. Steinhauser and Tuma~\cite{Steinhauser2020} proposed extending existing black-box scanners with a database response payload injection mechanism. For this sake, the authors designed a database traffic interception protocol that records the activity of web application databases and injects exploited patterns. Accordingly, the database reports to the analyzer every injection of a data string. Regular expressions are used to match injected patterns in received responses. The analysis process was much slower compared with ordinary black-box scanners.
	
	\item \textbf{Hybrid dynamic analysis:} hybrid dynamic analysis combines two or more dynamic analysis techniques for the detection of injection vulnerabilities. Two hybrid solutions have been found in the literature~\cite{Melicher2021, Ayeni2018}. Melicher et al.~\cite{Melicher2021} proposed and experimented the use of a deep learning model as a pre-filter for a dynamic taint-tracker. A deep learner model is used to check the vulnerability of JavaScript codes, only unconfirmed malicious samples are subject to a dynamic taint-tracker for analysis. A deep learner is trained on a collected dataset of confirmed and unconfirmed malicious scripts. Dynamic taint-tracking is performed making use of a modified browser with embedded taint-track analyzer. The solution is browser dependent and produces high false positives.
	Ayeni et al.~\cite{Ayeni2018} proposed a black-box scanner with embedded fuzzy inference system. A vulnerability is detected if one of seven DOM-based API features contains unsafe data given as input in a URL. The values of DOM features are extracted form HTTP responses and subjected to a fuzzification process. 21 rules are used by the inference engine to produce a fuzzified output which is deffuzified and interpreted as a presence or absence of a vulnerability. More experiments are needed to confirm the suitability of the proposal with large web applications.
	
\end{itemize}

\paragraph{\textbf{Test-case generation}}
Some static analysis based approaches, such as taint-analysis, and all dynamic analysis based approaches require a collection of input data to perform their assessment processes. The collection should include various, distinct and representative input tests to cover all the execution paths or scenarios and reveal potential fails. For systematic generation of input tests, several techniques are explored:

\begin{itemize}
	\item\textbf{Combinatorial security testing:} test inputs are designed as a model named input parameter model (IPM). The model is described as a context free grammar constituted of a finite number of parameters, each of which can take one of a finite set of values. A valid input test is obtained through  \textit{t-wise} combinations of input parameter values with a specific value \textit{t}. Constraints are useful means to reduce the input space and enforce the generation of valid, more comprehensive and sophisticated test inputs~\cite{Kuhn2013}. Six studies adopted such technique for the generation of successful XSS attack vectors. Table~\ref{tab:ct} describes the details of the models proposed in each study together with the targeted attacks. Specifically, Simos et al.~\cite{Simos2019} proposed three sub-grammars, each targets a specific HTML context.
	
	\begin{table}[h]
		\centering
		\small{
			\caption{Adopted combinatorial security testing models}
			\label{tab:ct}       
			\begin{tabular}{lrcrr}
				\hline\noalign{\smallskip}
				study & \#parameters & best \textit{t} value & \#generated cases& attacks\\
				\noalign{\smallskip}\hline\noalign{\smallskip}
				Bozic et al.~\cite{Bozic2015a} & 11 & 4 & 8761 & RXSS, SXSS\\
				Simos and Kleine~\cite{Simos2016} & 7  & 4 & 6891 & RXSS\\
				Garn et al.~\cite{Garn2019} & 7  & 3 & 7200 & RXSS\\
				Simos et al.~\cite{Simos2019} & 5, 4, 3  & 3 & 149, 114, 27& RXSS\\
				Garn et al.~\cite{Garn2021} & 12  & 2 & 99 & Unspecified\\
				Leithner et al.~\cite{Leithner2021} & 2  & 2 & 42 & RXSS\\
				\noalign{\smallskip}\hline
		\end{tabular}}
	\end{table}
	
	\item \textbf{Genetic algorithms:} in the aim to find the minimal number of test cases that reveal as many XSS vulnerabilities as possible, three studies combined static taint-analysis and genetic algorithms (GA). Firstly, a static taint-analyzer is used to identify vulnerable paths in the code of web applications. Then, a GA is used to generate test inputs to be injected and confirm the vulnerability of those paths. The algorithm starts with an initial collected set of candidate solutions, computes the fitness value of each candidate, selects candidates based on computed fitness values and generates new candidates by altering selected ones making use of crossover and mutation genetic operators. This white-box based technique enables understanding web application vulnerabilities before fixing them, but also inherits the limitations of static analysis (i.e., high false positives).
	Ahmed and Ali~\cite{Ahmed2016} and Marashdih et al.~\cite{Marashdih2017a} adopted a similar approach. Pixy tool~\cite{Jovanovic2006a} is used to extract vulnerable paths and a genetic algorithm is used to confirm their vulnerabilities. The only difference is that in~\cite{Marashdih2017a}, the process was optimized through removing infeasible paths from the test process. The proposed solution produces less false positives but infeasible paths are require to be removed manually. Avancini and Ceccato~\cite{Avancini2011} proposed a new improvement where concrete symbolic execution~\cite{James1976} is used to avoid local optimum caused by the GA. Specifically, a local search based on constraint solver is used to minimize the local optimum effect of the GA by the selection of appropriate input values that traverse more target paths. This approach is repeated whenever a local optimum is found by the GA. 
	
	\item\textbf{Machine learning:} machine learning are also used to generate valid attack vectors for the detection of XSS vulnerabilities. Caturano~ et al.~\cite{Caturano2021} proposed a reinforcement learning based approach. An attack string is divided into five distinct sections. A reinforcement learning agent acts by modifying only one of those sections at each step. It is trained with the help of the user to reach a valid attack string using standard Q-learning~\cite{Watkins1992}. This approach has higher accuracy and low false positives compared with other automated scanners but requires an extensive developer intervention in the training process. Frempong et al.~\cite{Frempong2021} proposed a tool that uses machine translations (encoder/decoder) with POS tagging to generate JavaScript exploits from intents specified in natural languages. After training, the model becomes capable of generating executable benign and malicious samples to be tested in real-world applications. However, the model generates only exploits related to specified intents.
	
	\item\textbf{Rule based generation:} rules enforced by taint-tracking are also used for the dynamic generation of test-cases. Lekies et al.~\cite{Lekies2013} modified the Chromium open source browser by integrating byte-level taint-tracking mechanism into its embedded components. The taint-tracker identifies source, sink and applied built-in filters to each tainted value in data flows. Those information are used by a control backend to generate valid XSS exploits appended to URLs to avoid embedded filtering capabilities of browsers. The generation is performed making use of well-defined rules associated to each source (HTML tags, nodes, comments or Javascipt). The technique is fully automated and enables the generation of exploits for the detection of DXSS vulnerabilities. Bensalim et al.~\cite{Bensalim2021} improved the generation rules adopted in~\cite{Lekies2013} by specifying URL positions where injected exploits are most likely to result in successful attacks. This enabled the detection of 1.9\% more vulnerabilities. The approach is limited to direct paths from sources to sinks with disabled URL-encoding which makes it less effective against complex attacks and real-world scenarios.
	Wang et al.~\cite{Wang2018} proposed a system that receives URLs and preprocesses them, loads their correspondent pages and obtains taint-traces. Those are used to automatically generate attack vectors through well-defined rules and verify their effects. The approach enabled the detection of 1.8\% more vulnerabilities than the commercial AWVS vulnerability scanner\footnote{Acunetix web vulnerability scanner: \url{https://www.acunetix.com/plp/web-vulnerability-scanner/}}. However, the technique is time expensive when it comes to generate complex attack vectors.  
	
	\item\textbf{Hybrid based test-case generation:} Kiezun et al.~\cite{Kiezun2009} proposed a hybrid and automatic generation technique of executable input tests for both XSS and SQL injection vulnerabilities. The process starts with automatically generated inputs by Apollo~\cite{Shay2008}, an input generator based on concrete and symbolic execution~\cite{James1976}. Those inputs are injected into web applications and checked if they reach sensitive sinks. If it is the case, inputs are mutated making use of an attack pattern library. Besides RXSS, the technique enables the detection of SXSS vulnerabilities through tracking flows of tainted data in web application databases. However, the proposed technique does not consider attacks across multiple pages and it is language based solution (PHP/SQL web applications).
	
	\item \textbf{Benchmarking:} benchmarks of test-cases are indispensable for the development of robust models and fair evaluation of different XSS detection methods.
	Pan and Mao~\cite{Pan2016a} proposed a micro-benchmark containing 175 test cases. Those test-cases are automatically generated from a template. The template is an abstraction of a typical vulnerable HTML page to DXSS attacks, it is constituted of six elements: source, propagation, transformation, sink, trigger and context. Associating values to each of those properties resulted on an executable attack vector. Enlarging the values of those properties enables the generation of more test-cases. However, the proposed benchmark does not cover complex attacks caused by language features and browser quirks.
	
\end{itemize}

\paragraph{\textbf{Hybrid analysis}} Hybrid analysis consists of using static and dynamic analysis in combination or alternation. Besides the detection of all data-flow paths, it is intended to produce low false positive rates.
Pan and Mao~\cite{Pan2017} proposed an alternation of static and dynamic analysis for the detection of DXSS vulnerabilities caused by the Greasemonkey browser extension~\cite{Kristen2008}. Greasemonkey is a cross platform extension that enables users to write personalized JavaScripts to customize web page appearances and behaviors. In the static analysis phase, user scripts from of the Greasemonkey extension are filtered by matching patterns related to specific privilege granting directives. In addition, a static parser is used to identify source-sink paths and enable the elimination of safe contents. Remaining scripts are subjected to a dynamic analysis (i.e., symbolic execution) to confirm their vulnerabilities. The approach has large overhead and requires manual assistance for the generation of suspicious event sequences.
Van Acker et al.~\cite{Van-Acker2012} used a combination of static and dynamic analysis for the detection of rich Internet application vulnerabilities to XSS. Static analysis is used to automatically identify the set of ActionScript sensitive variables used in files that can be affected with user inputs. This step is performed through de-compiling SWF files and using regular expression matching. Dynamic analysis is used to test identified variables in the aim to discover vulnerabilities. Malicious payloads are generated making use of 10 templates and injected as values to sensitive variables and checked for malicious effects. The approach fails to cover all vulnerabilities due to the adoption of an automatic clicker simulator which resulted on high portion of false positives.  
Balzarotti et al.~\cite{Balzarotti2008} combines static and dynamic analysis for testing the effectiveness of used sanitization. Static taint-analysis is used to identify applied sanitizers in web applications. Dynamic analysis is used to check the effectiveness of sanitizers by injection and tracking malicious inputs. If sensitive sinks are reached, sanitizers in the correspondent path are reported as ineffective.

\subsubsection{XSS vulnerability defense techniques} In order to protect web applications from XSS vulnerabilities at runtime, two major techniques are proposed: \textit{runtime data-code separation} and \textit{input parameter type inference and validation}. 

\paragraph{\textbf{Runtime data-code separation}} In order to prevent the execution of malicious scripts already injected into vulnerable web applications, scripts at sensitive points of HTTP responses can be extracted, at the server-side, and sent separately in a safe manner to the client-side in response to each HTTP request. If handled appropriately, malicious scripts can be prevented from being executed by web browsers.
In this context, Louw and Venkatakrishnan~\cite{Louw2009} proposed a code-data separation method that guarantees the safe construction of HTML parse trees on the web browser. Special codes are instrumented into sensitive points of web applications to call the server whenever reached. In response to each call, the parse tree of data at each point is prepared at the server-side and transmitted as string literals to prevent its execution by the browser. This releases browsers from handling suspicious scripts. In this proposal, developers are asked to manually annotate code positions holding untrusted data which is impractical for large web applications.
Inspired from SQL binding mechanism, Iha and Doi~\cite{Iha2009} proposed a similar technique for XSS. Web applications and browsers are modified to communicate a user modified agent request header. HTTP responses will then include the HTML structure without any parameter value. Those are included, following a specified syntax, in the HTTP response header. Consequently, the browser generates the DOM tree with empty parameter values and then uses the binding values from the HTTP response header and replaces them as literals that cannot be executed by browsers. Unfortunately, the proposed scheme does not support complex HTML structures, events and style attributes.
Parameshwaran et al.~\cite{Parameshwaran2015} used dynamic taint-analysis to infer benign DOM tree templates that can be generated with the presence of malicious user inputs. Those templates are stored into a database which is used at runtime to decide the best template fitting the runtime input. The experiment showed that the solution has a reasonable overhead but requires major changes in the code of web applications as well as browsers.
Gupta et al.~\cite{Gupta2019a} proposed a different approach. For every HTTP request, the server generates the correspondent HTTP response, extracts and isolates scripts in separate files and modifies the response code accordingly. At the client-side, extracted scripts are analyzed using taint-analysis and untrusted variables are identified. The output string of each untrusted variable is checked. If no suspicious pattern matches, the HTTP response continues to the user, otherwise the parameter value of the HTTP request and the URI link, if any, are extracted and decoded. A similarity with stored suspicious variables is measured. If similarity found, a vulnerability is reported and users are redirected to safe or error pages. The authors showed that the approach is convincing in terms of detection of vulnerabilities but no performance overhead analysis is performed. 

\paragraph{\textbf{Input parameter type inference and validation}} Most scripting languages used for developing complex web applications are untyped languages. This is exploited by attackers to inject scripts instead of clear values to input fields which leads to XSS attacks. Therefore, enforcing types on sensitive data values is an intuitive solution to protect against XSS attacks. In this context, Scholte et al.~\cite{Scholte2012} proposed a system that infers and validates input parameter values to protect against XSS and SQL injection vulnerabilities. At the training phase, types of input parameter values are inferred from benign test inputs and stored into a database. At runtime, types of input parameter values included in HTTP requests are inferred. Received and stored types are compared; if no match is detected, HTTP requests are dropped. The technique is fully automated but it is unable to extract all possible parameters especially in the case of encrypted or encoded HTTP requests.

\subsubsection{XSS Vulnerability detection and removal techniques}
For complete protection, studies proposed detection and repair techniques.
Shar and Tan~\cite{Shar2012a} used static taint-analysis from their previous work~\cite{Shar2012} to identify input and potentially vulnerable output nodes from the generated control-flow graphs of web application pages. Pattern matching is used to identify the HTML context of each node. Finally, adequate escaping mechanisms of OWASP are identified by matching the OWASP XSS prevention rules and instrumented making use of the ESAPI API\footnote{OWASP ESAPI API: \url{https://owasp.org/www-project-enterprise-security-api/}}.
Marashdih and Zaaba~\cite{Marashdih2017} proposed an approach for the detection and removal of RXSS and SXSS vulnerabilities from PHP web applications. Static taint-analysis of Pixy tool~\cite{Jovanovic2006a} is used to examine the source code of web application pages and generate their correspondent control-flow graphs. For optimization purpose, infeasible paths are removed from the resulted graphs. A GA is used to generate different test-cases from an initial population of XSS attack vectors. The fitness function is used to evaluate the traverse of each input to target paths; a path is considered vulnerable if the GA generator succeeds to traverse it with 0 fitness value. For removal, the HTMLPurifier library\footnote{HTML Purifier: \url{http://htmlpurifier.org}} is used to sanitize user data at detected vulnerable paths. The solution is language dependent and, like the proposal of Shar and Tan~\cite{Shar2012a}, it is unable to repair vulnerabilities caused by information flows across multiple pages.

\subsection{XSS attack detection techniques}
\label{sec:detection}
Runtime detection of attacks forms the last defense line against new and unknown attacks in the web. It is needed to cover risks that vulnerabilities' analysis and repair fail to deal with. It can also be considered as an alternative protection solution when detection and repair of vulnerabilities become hard to be performed in certain circumstances such as the case of legacy, large and critical web applications. The techniques of this kind can be installed independently of web applications at the client, server, proxy or multiple sides to detect and report attacks when they occur. The present review identified two explored techniques for runtime attack detection: \textit{HTTP traffic monitoring} and \textit{machine learning}.  

\subsubsection{HTTP traffic monitoring} By runtime HTTP traffic monitoring, every request and response related to a running web application is captured and checked for potential alteration by a non-controlled user data. Five studies adopted such approach are summarized in Table~\ref{tab:vuldetmonito}. Each solution proposal is described in terms of adopted detection scheme, installation location, pros and cons. 

\subsubsection{Machine learning} Different machine learning based models have been built for the detection of XSS attacks. Those models are developed in the aim to learn hidden properties of attack vectors and make correct predictions at runtime. Twenty-three studies included in this review used machine learning to achieve this aim. Different types of machine learning have been explored: single, ensemble and deep learning. Table~\ref{tab:attdetml} summarizes the proposed models together with the type and number of features and the best obtained performance for each model. In the sequel we discuss only remarkable proposals:

Besides traditional features extracted from URL and HTML contents, Wang et al.~\cite{Wang2014} and Rathore et al.~\cite{Rathore2017} experimented new kind of features named OSN for the detection of XSS worms (WXSS). The OSN features capture the observed behavior on targeted social networking services. They include the spreading speed and frequency of suspicious data in the network traffic. The authors experimented several classifiers with and without the OSN features and found that the new features are more discriminating.

Mereani and  Howe~\cite{Mereani2018a} proposed a two stage based classification system. At the first stage, a decision tree classifier is used to predict the nature of user inputs (i;e., texts or scripts). Scripts are subjected to a second classification stage with an ensemble classifier to predict their maliciousness. In total, 62 alphanumeric and non-alphanumeric features are extracted and used for classification. 

Zhang et al.~\cite{Zhang2019} used two separate Gaussian mixture models (GMM) trained on normal and XSS payloads respectively. The produced probabilities of the two GMMs on tested payloads are compared to reach a final prediction. They found that the combination of features extracted from HTTP requests and responses increases the accuracy score of the model.

Li et al.~\cite{Li2020} used Random Forest (RF) as a semi-supervised learner. To rectify mislabeled data, a weighted neighboring process (weighted KNN) is adopted to re-label samples regarding their weighted distances to other similar samples. 

Zhou and Wang~\cite{Zhou2019} used an ensemble learner formed by a voting of several Bayesian Networks. In the aim to improve the prediction of the proposed model, the predictions obtained from the network are combined with a set of extracted threat intelligence rules including malicious IP addresses and domains from PhishTank\footnote{PhishTank: \url{https://phishtank.org}} and FireHOL\footnote{FireHOL: \url{https://firehol.org}}.

In order to enforce the detection of machine learning based models against adversarial attacks, several evolutionary techniques are proposed for the generation of adversarial attacks from regular attacks. Most of these techniques used reinforcement learning to achieve this goal:

Fang et al.~\cite{Fang2019} proposed a system that incorporates a detection model and an adversarial attack model. XSS payloads detected by the detection model are subjected to a set of disguising actions performed by an agent based on a Double Deep Q-Network algorithm (DDQN). Those actions include encoding, obfuscation, sensitive word substitution, position morphology transformation and adding special characters. Modified samples are transmitted back to the detection model. If a modified sample is predicted as benign, it is labeled as malicious and used for retraining the adversarial and prediction models; otherwise, the sample is sent back to the agent for further deforming. The process ends when a number of attempts is reached based on a measured feedback reward formula.

Zhang et al.~\cite{Zhang2020} used an adapted Monte Carlo tree search algorithm (MCTS) for the generation of XSS attacks. Adversarial attacks are generated following a set of bypassing rules enabling hexadecimal encoding, decimal encoding, url encoding, insertion of invalid chars in the middle of tags and case mixture.

Wang et al.~\cite{Wang2022} proposed a system that starts by applying fuzzing to build a dataset of malicious and benign samples. Those samples are used as inputs to an adversarial attack model based on reinforcement learning. Two agents based on Soft-Q learning algorithm are proposed for the generation of adversarial attack samples. The former applies escaping actions on HTML tags while the latter performs actions on JavaScripts. Those actions include: string substitution, char coding and string addition.

Tariq et al.~\cite{Tariq2021} proposed a mixed approach for the detection of XSS attacks based on genetic algorithm, statistical inference and reinforcement learning. The aim was to improve the detection of new XSS attacks. The proposed approach uses a genetic algorithm to check the distance of each payload to malicious and benign samples. If the algorithm fails due to overlaps, a statistical inference module is used. In addition, a threat intelligence (IP addresses and malicious domains) is used to enforce the accuracy of the prediction. A payload is considered safe only if the three detection models predict it as safe, otherwise the payload is considered malicious. Reinforcement learning is used to add and update the samples to recognize more new payload attacks. The proposed approach reached 99.89\% of accuracy.

For banchmarking, Mokbal et al.~\cite{Mokbal2020} proposed an algorithm named C-WGAN-GP for oversampling datasets without overfitting effects. The proposed algorithm generates synthetic but valid and reliable samples of the minority class. They are also indistinguishable from real XSS payloads.

\subsection{XSS attack defense techniques}
\label{sec:defense}
Dynamic defense techniques provide mechanisms to protect systems against realtime attacks, block the execution of malicious actions embedded on payloads and prohibit their propagation. The study of the literature distinguishes three main mechanisms: \textit{filtering}, \textit{moving target defense} and \textit{sandboxing}.

\begin{landscape}
		\begin{longtable}{p{2.9cm}p{5.5cm}p{1cm}p{1cm}p{4cm}p{5.5cm}}
			\caption{XSS attack detection through HTTP traffic monitoring}
			\label{tab:vuldetmonito}\\      
			\hline\noalign{\smallskip}
			study & detection scheme & location & attacks & pros & cons\\
			\noalign{\smallskip}\hline\noalign{\smallskip}
			Johns et al.~\cite{Johns2008} & For RXSS, they match scripts extracted from HTML requests and responses after removing all encoding and performing a threshold-based similarity match. For SXSS, a model is trained on scripts included in the web-application after the execution under benign data and any detected deviation is reported as an attack. & server-side & RXSS, SXSS &
			low false positive rates (0.5\% for RXSS and 0.7\% for SXSS)
			& 
			\begin{minipage} [t] {0.34\textwidth} 
				\begin{itemize}
					\item SXSS detection relies upon a trained model, since it is impossible to cover all the input scenarios, the solution may has higher false positives for more complex web applications.
					\item Relies on a strong assumption that the attacker cannot add external scripts to trusted domains.
				\end{itemize} 
			\end{minipage}\\
			
			\noalign{\smallskip}\hline\noalign{\smallskip}
			Sun et al.~\cite{Sun2009} & Extracted scripts from HTTP requests are analyzed, recursively decoded and stored. DOM trees of HTTP responses are generated and scripts are extracted from sensitive locations. A similarity is checked between decoded scripts resulting from HTTP requests and responses. A match is reported as an XSS attack. & client-side & WXSS &
			effective for the detection of self-replicating XSS worms on the client side with reasonably low performance overhead (0.78\%-6.75\%).
			& 
			\begin{minipage} [t] {0.34\textwidth} 
				\begin{itemize}
					\item not suitable for server-side self-replicating worms. 
					\item ineffective with the presence of customized obfuscation techniques.
				\end{itemize} 
			\end{minipage}\\
			
			\noalign{\smallskip}\hline\noalign{\smallskip}
			Sundareswaran and Squicciarini~\cite{Sundareswaran2012} & Comparison of generated and stored control flow graph signatures of web pages at the proxy server. Features are extracted from both versions and a threshold value is used to check for potential similarities. A deviation is reported as an attack. & client and proxy& RXSS, SXSS, DXSS &
			low performance overhead (less than 1\%).
			& 
			\begin{minipage} [t] {0.34\textwidth} 
				\begin{itemize}
					\item quite higher false positive rate (3.75\%).
					\item relies on the presence of clean pages in the server.
				\end{itemize} 
			\end{minipage}\\
			
			\noalign{\smallskip}\hline\noalign{\smallskip}
			Das et al.~\cite{Das2013} & Matching runtime execution-sequences to stored legitimate sequences at the training phase. Any deviation is reported as an attack.  The legitimate execution sequences list is updated by the administrator when unsuccessful communications are reported by clients. & client-side &
			RXSS, SXSS, DXSS & satisfactory results obtained for different XSS attacks.
			& 
			\begin{minipage} [t] {0.34\textwidth} 
				\begin{itemize}
					\item the automatic update of execution sequences lists is a challenging task. 
					\item requires further experiments with real-wold applications.
				\end{itemize} 
			\end{minipage}\\
			
			\noalign{\smallskip}\hline\noalign{\smallskip}
			Yamazaki et al.~\cite{Yamazaki2018} & Matching generated and stored templates identified by a restoration algorithm. Any match is reported as an attack. & proxy-side & RXSS, SXSS &
			overhead varies from moderate to low.
			& 
			\begin{minipage} [t] {0.34\textwidth} 
				\begin{itemize}
					\item higher false positive rate (20.6\%). 
					\item relies on the presence of clean pages at the server.
				\end{itemize} 
			\end{minipage}\\
			
			\noalign{\smallskip}\hline
		\end{longtable}
\end{landscape}

\begin{landscape}
		\begin{longtable}{lllll}
			\caption{XSS attacks detection through machine learning}
			\label{tab:attdetml}\\      
			\hline\noalign{\smallskip}
			study & type & features (\#) & targeted attacks & best perf.\\
			\noalign{\smallskip}\hline\noalign{\smallskip}
			\multicolumn{4}{c}{Single learning}\\
			\noalign{\smallskip}\hline\noalign{\smallskip}
			Nunan et al.~\cite{Nunan2012} & SVM & URL and HTML (6) & RXSS, SXSS, DXSS & ACC = 99.89\% \\
			
			\noalign{\smallskip}\hline\noalign{\smallskip}
			Goswami et al.~\cite{Goswami2017} & kmeans & JavaScript (16)  & RXSS & ACC = 98.89\% \\
			
			\noalign{\smallskip}\hline\noalign{\smallskip}
			Mokbal et al.~\cite{Mokbal2019} & MLP & URL and HTML (41)  & Unspecified  & ACC = 99.32\% \\
			
			\noalign{\smallskip}\hline\noalign{\smallskip}
			Mereani and Howe~\cite{Mereani2021} & KNN & scripts from HTTP requests (16)  & Unspecified & ACC = 99.86\%\\
			
			\noalign{\smallskip}\hline\noalign{\smallskip}
			Mokbal et al.~\cite{Mokbal2022} & SVM & vectorized payloads (96) & Unspecified & F1 = 99.59\% \\
			
			\noalign{\smallskip}\hline\noalign{\smallskip}
			\multicolumn{4}{c}{Ensemble learning}\\		
			\noalign{\smallskip}\hline\noalign{\smallskip}
			Wang et al.~\cite{Wang2014} & AdaBoost & URL, HTML and OSN features (12) & WXSS & F1 = 93.90\%\\
			
			\noalign{\smallskip}\hline\noalign{\smallskip}
			Rathore et al.~\cite{Rathore2017} & RF & URL, HTML and OSN features (25) & WXSS & F1 = 97.40\% \\
			
			\noalign{\smallskip}\hline\noalign{\smallskip}
			Mereani and Howe~\cite{Mereani2018a} & Stack(SVM-L, SVM-P, RF, NN) & scripts from HTTP requests (62)  & SXSS &ACC = 99.97\% \\
			
			\noalign{\smallskip}\hline\noalign{\smallskip}
			Zhou and Wang~\cite{Zhou2019} & Vote(xBN) + Threat intelligence rules & scripts from HTTP requests (30) &Unspecified & ACC = 98.54\% \\

			\noalign{\smallskip}\hline\noalign{\smallskip}
			Zhang et al.~\cite{Zhang2019} & Stack(2GMM) & vectorized HTTP requests/responses (200)  & RXSS, SXSS & ACC = 96.49\% \\
			
			\noalign{\smallskip}\hline\noalign{\smallskip}
			Nagarjun and Ahamad~\cite{Nagarjun2020} & HGBC & vectorized payloads (128)  & RXXS, SXSS, DXSS & ACC = 99.89\% \\
			
			\noalign{\smallskip}\hline\noalign{\smallskip}
			Li et al.~\cite{Li2020} & Semi-supervised RF + weighted-KNN & URL (12)  & Unspecified & ACC = 97.30\% \\
			
			\noalign{\smallskip}\hline\noalign{\smallskip}
			Malviya et al.~\cite{Malviya2021} & RF & scripts and HTML (44)  & Unspecified & ACC = 100\% \\
			
			\noalign{\smallskip}\hline\noalign{\smallskip}
			Mokbal et al.~\cite{Mokbal2021} & XGBoost & URL, scripts and HTML (160 reduced to 30)  & Unspecified & F1 = 99.58\%\\
			
			\noalign{\smallskip}\hline\noalign{\smallskip}
			\multicolumn{4}{c}{Deep learning}\\		
			\noalign{\smallskip}\hline\noalign{\smallskip}
			Kadhim and Gaata~\cite{Kadhim2020} & CNN with LSTM layer & vectorized payloads (3000)  & Unspecified & F1 = 99.30\%\\
			
			\noalign{\smallskip}\hline\noalign{\smallskip}
			Fang et al.~\cite{Fang2020} & Bi-RNN with attention mechanism & vectorized payloads (200)  & SXSS & F1 = 99.00\%\\
			
			\noalign{\smallskip}\hline\noalign{\smallskip}
			Chaudhary et al.~\cite{Chaudhary2021} & CNN & Tagged payloads + PCA (100) & RXSS, SXSS & F1 = 99.37\%\\
			
			\noalign{\smallskip}\hline\noalign{\smallskip}
			Liu et al.~\cite{Liu2022} & GCN = Graph Convolutional Network & vectorized payloads (200) & RXSS, SXSS & F1 = 99.70\%\\	
			
			\noalign{\smallskip}\hline
		\end{longtable}
\end{landscape}

\subsubsection{Filtering} Filtering is a blocking based technique. It is a simple way to defend against cyberattacks including XSS. Through filtering, distinguished XSS attacks from benign scripts are blocked automatically. Several filtering based approaches are found in the literature, they only differ on the way to distinguish XSS attacks.

\paragraph{\textbf{Whitelist/Blacklist based filtering}} Through identifying and listing all benign/malicious scripts, XSS attacks can be distinguished and blocked by the system. Actually, this is an infeasible solution since listing all malicious or benign scripts cannot be established in practice. To alleviate this problem, all scripts are considered malicious except those used by developers to construct the target web applications.

Wurzinger et al.~\cite{Wurzinger2009} proposed a proxy based filtering technique. All scripts used by developers in web applications are encoded and every HTTP response is received and sent to a modified browser in the proxy server. This later checks for any clear script. Those scripts are identified as malicious and blocked. If no clear script is generated, original encoded scripts are decoded and clear HTTP responses are safely delivered to users. The solution introduces a large overhead due to frequent decoding of scripts at runtime and thereby not suitable for high performance web services.

Mitropoulos et al.~\cite{Mitropoulos2016} proposed a client-side filtering approach. The JavaScript engine of browsers is extended with a transparent script interception layer that identifies every script not in the list of valid scripts previously fixed by developers for each page; scripts not on the list are simply blocked. To avoid false positives caused by slight changes and dynamic code generation, the layer uses fingerprints previously created on the server-side when comparing scripts. For robustness, when script elements are altered or new scripts are added on the server side, a new fingerprint generation phase is required. Scripts not matching any fingerprint are blocked and an alert is sent to the administrator. The solution causes a negligible overhead (less than 0.05\%) but the initial and effective creation of fingerprints using dynamic analysis is hard. In addition, the proposed solution fails to detect all mimicry attacks (JSM-XSS) and requires the modification of browsers to integrate the script interception layer. 

Gupta and Gupta~\cite{Gupta2018c} used a database of  attack vectors associated to each injection point. These are identified by a hybrid analysis: static analysis is performed through parsing files to identify all the injection points and dynamic analysis is performed through injection of malicious attacks and observing the resulted behavior. At runtime, HTTP requests with injected payloads are analyzed by the server. Any match is reported as an attack and the requests are blocked. The database is updated whenever new attack vectors are detected. The experiment showed that the solution suffers from higher false positive and negative rates.

Chaudhary et al.~\cite{Chaudhary2020} proposed a proxy based filtering approach that uses a blacklist of XSS attacks. HTTP responses are transformed in a way that vulnerable scripts are isolated and saved in a separate file. They are next decoded and grouped following the Levenshtein distance similarity to reduce the refining phase. Finally, they are matched to XSS attack vectors stored in a repository. Matched scripts are sanitized making use of XSS filtering APIs and safe responses are delivered to users. Unfortunately, the solution depends on an attack repository which prevents the detection of all new attacks even with the frequent update if such repository.

Pazos et al.~\cite{Pazos2021} proposed a client-side solution that makes use of well-identified vulnerabilities recognized by analysts deployed in a form of well-formed signatures. A browser extension receives  HTTP responses, matches injection points and apply proper sanitizers matching signatures and injection points from already known vulnerability reports. The system maintains a database of such signatures for reuse. The aim is to reduce time from zero day attack to the deliverance of patches where they are directly applied by the browser. The solution causes a considerable overhead (10\%-50\%), depends on delivered signatures and does not support complex sanitizers.

\paragraph{\textbf{Regular expression based filtering}} Instead of explicitly set the list of scripts to be allowed or blocked, regular expressions are used to describe general patterns of such scripts. Those patterns are used by client or server-side filters to match and block malicious scripts at runtime. 

To defend against CS-XSS, Gebre et al.~\cite{Gebre2010} proposed a server-side filter making use of regular expressions modeling dangerous scripts that can be injected in specific HTML elements. The proposed filter has low overhead (60 ms) and produces no false positive but it is unable to distinguish benign HTML tags in PDF files. Moreover, designed regular expressions do not cover encoded and obfuscated scripts. 

Javed and Schwenk~\cite{Javed2014a} proposed a regular expression based filter to defend against XSS attacks on mobile based web applications. They improved the set of regular expressions defined in~\cite{Wassermann2008} for vulnerability detection and added new ones. In total, twenty-five patterns are defined. The filter is implemented as a JavaScript function embedded on the code of mobile web applications to filter inputs at runtime. The proposed filter is not suitable for desktop based web applications due to their complex nature and the significant use of AJAX based interactions. 

\paragraph{\textbf{Rule based filtering}} Rules can be used to match suspicious links and detect anomalies in input data. 

Shanmugam and Ponnavaikko~\cite{Shanmugam2007a} proposed a solution based on adding requirements for data inputs (type, size, allow special char, allow tags). Those requirements are transformed by a tool to XML schema and stored in a database. Therefore, when an HTTP request is received, the input values are extracted and converted into XML objects and then mapped to the correspondent stored schema. If there is a match, the HTTP request is forwarded to the application otherwise an attack is detected and blocked and the user is redirected to an error page. The solution is platform and language independent but requires tough modifications of the server for their integration.

Again, Shanmugam and Ponnavaikko~\cite{Shanmugam2007b} proposed another solution based on adding security attributes to each web application stored in the server. Those attributes identify the security level required for the application such as the maximum number of input chars, the encoding mechanism, and character-set. Those attributes are used to detect any anomaly behavior. The proposed system starts when receiving an HTTP request, the application attributes are loaded and the input is sanitized accordingly. If the input length exceeds the specified maximum number, the input is rejected; otherwise, if the input does not contain any specific char, the call is forwarded to the application, otherwise the input is parsed by separating it into tokens, the resulted vector is verified by checking a list of vulnerability identification rules that makes use of whitelist tags and their attribute values, if one of the rules is not satisfied, the user is redirected to an error page. The time to handle a single vulnerable input is low (0.04ms). However, the solution does not support all encoding patterns and for robustness, the list of whitelist tags should frequently be updated to follow the development of web programming technologies. 

Kirda et al.~\cite{Kirda2009} proposed an integration of a personal firewall that enables users to add new rules, update or remove existing rules. Rules are used to match risky links. Local links are ignored but cross-domain links are detected and users are notified if a suspicious link does not match any rule. Rules are removed through a garbage collector after being un-matched for a long period of time. Moreover, a limited number of external links is permitted otherwise no link is allowed, this is adopted to prevent binary-encoding attacks. The major flaw of the proposed solution is the dependence on users to update rules.

\paragraph{\textbf{Content-Security Policy (CSP)}} CSP is another type of filtering capabilities. It is a computer security standard that is actually fully or partially supported by most modern browsers. It enables the restriction of resources that the browser can load and execute. A CSP is specified at the server-side, transmitted as HTTP header in HTTP responses and executed at the client-side by browsers. CSPs are commonly used to defend against XSS attacks but are actually used to defend several code injection attacks.  

Jim et al.~\cite{Jim2007} proposed a basic content-security policy named BEEP. BEEP scheme enables the specification of a whitelist of hashed legitimate scripts identified by web page developers. Therefore, all scripts included in HTTP responses are detected prior rendering, hashed and matched to the whitelist of scripts. As a result, only matched scripts are executed. In addition, the policy enables DOM sandboxing making use of \textit{noexecute} nodes to prevent the execution of scripts potentially injected to \texttt{div} and \texttt{span} tags. The proposed solution is simple to implement and generates no false positives. However, the current implementation requires the manual identification of whitelist scripts by developers and causes a quite large overhead (14.4\% for whitelist policies and 6.6\% for DOM sandboxing). Johns~\cite{Johns2014} conducted an empirical study showing the limitation of CSP 1.1 in defending against JSM-XSS attacks and compromising whitelisted domains. To overcome such loopholes on the use of CSP, the author proposed a system named PreparedJS. The proposal is based on templating scripts and cryptographic script checksums. By templating scripts, placeholders where required data needed to be injected are marked with a specific syntax. Each placeholder is stored together with a list of allowed values in a JSON format. By cryptographic script checksums, prepared templates are hashed, so that every script received at the browser is hashed and matched to allowed scripts. The list of allowed script checksums are included in the security policy. The proposal has low overhead (54ms-148.6ms) but requires considerable modifications of web applications.

Stamm et al.~\cite{Stamm2010} introduced another basic CSP to protect web applications against XSS attack. The proposed scheme enables blocking inline scripts and prevents the transformation of strings into codes by blocking calls to the \textit{eval()} function. The proposed solution requires the modification of web applications to export inline scripts to external files. To alleviate this issue, Doupé et al.~\cite{Doupe2013} proposed a similar but automatic system named DeDacota. The aim was to separate code from data of web application pages and enforce the browser, through a CSP, to block any execution of inline scripts. However, no formal proof of correctness is provided for the tool. 

Fazzini et al.~\cite{Fazzini2015} proposed a system that automatically generates CSPs for web applications. Starting by a dynamic training with the help of input tests and taint-analysis, the system learns trusted and untrusted parts of web pages making use of annotated DOM trees. The system infers a policy that blocks untrusted parts while the source code of the application is transformed to meet the inferred policies. The transformations include moving inline scripts into external files and prevent inline scripts in the CSP. The proposed solution causes no significant overhead (less than 59 ms) but it may generate false positives that require manual intervention of developers. In addition, the solution cannot deal with source code statements that are dynamically generated by client-side codes.

Pan et al.~\cite{Pan2016} also proposed a system for the automatic generation of CSP policies for web applications. The proposed system is firstly trained for the inference of script templates. Secondly, web pages are rewritten so that benign and matched scripts of inferred templates are extracted and stored in a trusted subdomain and included as external scripts. This way only external scripts from the trusted subdomains are allowed. Dynamically generated scripts are matched to the templates before allowing their execution. If matched, the web page is modified on the fly to call the correspondent external file including the originally inlined script. This solution causes a moderate overhead (9.1\%) but suffers from the cold start problem.

To protect against DXSS, Iqbal et al.~\cite{Iqbal2019} proposed a CSP that prevents unauthorized alteration of DOM objects generated at the browser. The CSP policy is enforced by a DOM monitoring module installed in secure browsers. This module monitors client-side DOM behaviors and authorizes or blocks incoming requests. This enables a server-side control on the modifications performed on the DOM at the client-side. To prevent obfuscated requests, the DOM monitoring module deobfuscates HTTP requests before reaching the DOM-API. The proposed solution is resilient to obfuscated malicious requests and causes negligible overhead. However, it requires modifications of browsers and manual specification of policies.

Xu et al.~\cite{Xu2020} proposed a new CSP named JavaScript CSP (JSCSP) that uses refined rules to defend injections to specific tags or elements instead of whole pages. It also enables the automatic generation of rules. When the JSCSP is deployed, DOMs of received pages are blocked temporarily and safer DOMs are generated, the policy rules are checked and malicious scripts are removed, then the original DOM rendering is enabled. The evaluation showed that JSCSP is compatible with different browsers but depends on the availability of clean pages, otherwise already injected vectors cannot be detected. Moreover, JSCSP is compatible with static websites since dynamic ones allow the addition of new elements dynamically to the DOM. In addition, JSCSP generates high false positives.

Mui and Frankl~\cite{Mui2011} proposed a quite different solution. They used two different encoding for web applications and user inputs. Trusted web application code is encoded with standard characters where user inputs are encoded using a complementary character encoding. This way, user inputs are easily tracked. Token-based security policies are used to prevent sensitive sinks being reached with special user tokens. This solution replaces the need for introducing sanitizers and causes low overhead (0.17\%-1.74\%). However, it requires the modification of the server and browser components. It also requires the modification of web applications to remove sanitizers that may cause conflicts with the proposed scheme.

\paragraph{\textbf{Runtime taint-tracking}} Runtime taint-tracking enables controlling whether attacker injected data in URLs reach sensitive sinks and cause the browser to execute their incorporated scripts. 

Vogt et al.~\cite{Vogt2007} used runtime taint-tracking to prevent user sensitive data from being sent to third-parties. The JavaScript engine of the browser is modified to support dynamic (bytecode) taint propagation. Therefore, whenever a JavaScript program attempts to transmit sensitive data, the system checks whether the host of the loaded page and the host to which sensitive data is sent are from the same domain. If it is not the case, users are alerted to take the appropriate actions. This is only suitable for experienced users.

Stock et al.~\cite{Stock2014} used runtime-taint tracking to stop parsing scripts other than those used by trusted developers. For this sake, the JavaScript engine is modified to enable the tokenization of scripts, the identification of script origins and the ignorance of any code containing non-literals. The rendering engine accepts only tainted data to appear in the protocol and the domain of remote URLs from the same origin. In addition, an API is provided to specify when the application should activate the dynamic generation of elements to avoid blocking desirable behaviors. The solution causes a moderate overhead (between 7\% and 17\%) but requires extensive modifications of browsers.

\paragraph{\textbf{Upload filtering}} Upload filtering is a server-side technique that checks file contents for potential injection of malicious scripts when they are uploaded in the server. Suspicious detected files are either modified by removing injected scripts or simply rejected. It is a powerful technique to defend against CS-XSS. It is explored by two studies. Barth et al.~\cite{Barth2009} implemented an upload filter constituted of 34 HTML signatures. So that, any file content matching any of those signatures is reported as a suspicious file and blocked by the server. Only the initial bytes of files are inspected while scripts can be inserted anywhere. To overcome this limitation, Barua et al.~\cite{Barua2011} developed a two stage detection filter. Firstly, the MIME type is detected from file extensions and metadata. If the MIME type cannot be detected or it is not whitelisted, the upload is rejected. Secondly, files with known and benign MIME types are subjected to a content analysis. This later incorporates an encoding process and a content parsing for locating JavaScript method calls. The presence of JavaScript tags indicates the suspiciousness of files, thereby a browser emulator is invoked to determine any malicious actions when downloading. However, the solution causes large overhead (from 353\% to 3030\%).

\subsubsection{Moving Target Defense (MTD)} MTD techniques are designed to reduce the success rate of attacks by continuously shifting the configuration and parameter values of running applications. This hardens the understanding of structure and behavior of web applications which increases the uncertainty of adversaries and protects against traditional attacks. Randomness is the key success of MTD techniques since deterministic changes can be learned by attackers and hence permit them to design new successful attack vectors. MTD techniques has shown their ability to defend against sophisticated and complex cyber-attacks~\cite{Bradley2022}. For defending XSS attacks, eight studies are identified by this review:

Nadji et al.~\cite{Nadji2009} proposed a client-server solution that prevents altering the structure of trusted contents by untrusted data. At the server-side, trusted and user-generated data are separated by annotating trusted data using randomly generated markup primitives. At the browser-side, untrusted user-generated data can easily be distinguished at the parsing level, and then isolated and tracked dynamically. Server-side policies are used to confine untrusted data. This solution has low false positives, browser-level and server-level overheads (1.1\%-1.85\% at the browser and 1.2\%-3.1\% at the server). It requires  modifications of web applications, servers and browsers to support the parser level isolation and interpretation policy enforcement.

Athanasopoulos et al.~\cite{Athanasopoulos2010} proposed a system that automatically extract all JavaScripts of web pages, encrypt them and transpose them to a randomly generated domain. The decryption key is transmitted to the client side through action-based policies in the HTTP header. Therefore, attacker injected codes cannot bypass the policy. The evaluation process showed that the system has a negligible computational overhead but the integration of the system requires modifications on browsers.

Shariar and Zulkernine~\cite{Shahriar2011} proposed a similar approach to deterministic moving target defense techniques. Each stored web application file is inspected where injection points are firstly identified and comments are inserted before and after each point. Injected comments are stored with the expected features of the enclosing points (i.e., number of tags, number of attributes). For every received request, the server generates the response instrumented with injected boundaries, and the features are calculated and compared with the stored ones, each deviation indicates a potential XSS attack. In such case, malicious contents are removed. Injected boundaries are removed before the response is sent to the browser. The solution produces false positives that reach up to 5.2\% with a reasonable response-time delay that ranges from 2\% to 6\%. However, it is language dependent since appropriate parsers are required to identify every injected point. In addition, the deterministic nature of injected comments may cause attackers to learn the process and launch adversarial attacks. A similar but improved approach has been proposed later on by Gupta and Gupta~\cite{Gupta2016}; in this latter work, injected comments are formed from random generated strings with runtime sanitizer instrumentation for avoiding the execution of malicious scripts. However, the new proposal produces higher false positives that range from 10\% to 15\%. Later on, a cloud based version for OSN web applications is proposed by the same authors in~\cite{Gupta2017a, Gupta2020a}. 

Van-Gundy and Chen~\cite{Van-Gundy2012} proposed an MTD based solution that enables browsers to distinguish between benign and malicious contents of web pages. The server used a function to randomly generate strings used as prefixes to the (X)HTML tags in response of each HTTP request. This way, every injected content will be easily distinguished by the browser. Those contents are blocked making use of a security policy transferred by the server and checked by the browser before rendering the page. The solution proposal is simple and produces no false positives with moderate overhead (14\%-20\%). But, it requires server-side modifications and manual specifications of policies for each application.

Niakanlahiji and Jafarian~\cite{Niakanlahiji2019} proposed a different MTD technique. They modify web applications by adding a new attribute named \textit{runtimeID} to suspicious tags with randomly generated hex value. The head of each page is also modified to include a script that will be executed for every user request. The role of that script is to generate a random hex identifier and add it to each \textit{runtimeID} attribute in the page. At the browser side, new added elements without a \textit{runtimeID} attribute or with different \textit{runtimeID} value are considered as malicious and are removed automatically. The solution produces no false positives and has low average overhead (4ms). Unfortunately, it requires manual interventions to add the \textit{runtimeID} attribute to sensitive elements.

\subsubsection{Sandboxing/Isolation} Sandboxing is an isolation technique that enables the full control of running applications and prevents suspicious actions to be propagated. Cao et al.~\cite{Cao2012} proposed a refined sandboxing technique to prevent WXSS attacks. Rendered pages are isolated at the client-side. Every web page is divided into different and separate parts named views according to a specific strategy. Each view has a list of predefined authorized actions. Therefore, views cannot perform and request actions from other views without authorization. For isolation, pseudo-domains are associated to views and each view has a list of controlled capabilities. This way, every required connection from one view to another is controlled by an authentication mechanism. For example, a request to the server to modify a content-profile will be verified and blocked if it is originated from an infected page (not authorized action). The request emanating from a view that has no capability to send it, is blocked. This can be implemented by techniques like per-url session tokens and referrer-based view validation. The technique has a reasonable latency and memory overhead (about 1\% and 4\% respectively) but involves modifications of web applications to define views and their correspondent authorized actions.

\subsubsection{XSS attack detection and defense techniques} Nine studies included in this review proposed a runtime detection with defense techniques. Remarkably, runtime sanitizer instrumentation is the dominant approach adopted as a defense technique. Table~\ref{tab:athybrid} summarizes those studies highlighting their targeted attacks and limitations.

	\begin{longtable}{p{3.1cm}p{2.8cm}p{2.8cm}p{1cm}p{4.2cm}}
		\caption{XSS attack detection and defense techniques}
		\label{tab:athybrid}\\      
		\hline\noalign{\smallskip}
		study & detection method & defense method & attacks & limitations \\
		\noalign{\smallskip}\hline\noalign{\smallskip}
		Bisht and Venkatakrishnan~\cite{Bisht2008} & HTTP traffic monitoring and matching shadow and real generated pages. Shadow pages are obtained through transmitting HTTP requests with benign string inputs. & Remove all scripts not present in intended pages. & Unspecified &
		\begin{minipage} [t] {0.27\textwidth} 
			\begin{itemize}
				\item performance overheads ranged from 5\% to 24\%.
				\item produces false positives when attacks use non-Firefox quirks.  
			\end{itemize} 
		\end{minipage}\\
		
		\noalign{\smallskip}\hline\noalign{\smallskip}
		Bates et al.~\cite{Bates2010} & HTTP traffic monitoring and matching scripts from HTTP requests and responses after recursive decoding & Refuse to deliver matched scripts to the JS engine. & RXSS &
		\begin{minipage} [t] {0.27\textwidth} 
			\begin{itemize}
				\item limited to the detection of single injections omitting multiple injections.
				\item requires modifications on the browser.
			\end{itemize} 
		\end{minipage}\\
		
		\noalign{\smallskip}\hline\noalign{\smallskip}
		Pelizzi and Sekar~\cite{Pelizzi2012} & HTTP traffic monitoring and matching scripts from HTTP request and response parameters & Refuse to deliver matched scripts to the JS engine. & RXSS &
		\begin{minipage} [t] {0.27\textwidth} 
			\begin{itemize}
				\item requires the modification of the browser to check the execution permission of each script. 
				\item unable to detect scripts subjected to extensive string transformations by web applications.
			\end{itemize} 
		\end{minipage}\\
		
		\noalign{\smallskip}\hline\noalign{\smallskip}
		Gupta and Gupta~\cite{Gupta2016b1} & HTTP traffic monitoring and matching scripts from HTTP requests and DOM trees generated for HTTP responses & Runtime context-aware sanitizer instrumentation. & WXSS &
		\begin{minipage} [t] {0.27\textwidth} 
			\begin{itemize}
				\item low detection rate (F1-score less than 95\%).
			\end{itemize} 
		\end{minipage}\\
		
		\noalign{\smallskip}\hline\noalign{\smallskip}
		Gupta and Gupta~\cite{Gupta2016e} & HTTP traffic monitoring and matching scripts from HTTP requests and responses & Runtime context-aware sanitizer instrumentation. & RXSS, SXSS, DXSS &
		\begin{minipage} [t] {0.27\textwidth} 
			\begin{itemize}
				\item no performance overhead analysis is performed.
			\end{itemize} 
		\end{minipage}\\
		
		\noalign{\smallskip}\hline\noalign{\smallskip}
		Gupta et al.~\cite{Gupta2018d} & HTTP traffic monitoring and matching scripts stored and generated DOM trees & Runtime context-aware sanitizer instrumentation. & DXSS &
		\begin{minipage} [t] {0.27\textwidth} 
			\begin{itemize}
				\item quite large overhead due to runtime nested auto-context-sensitive sanitization.
			\end{itemize} 
		\end{minipage}\\
		
		\noalign{\smallskip}\hline\noalign{\smallskip}
		Lalia and Sarah~\cite{Lalia2018} & HTTP traffic monitoring and matching script features extracted form generated HTTP responses with those stored after performing a static analysis. & Runtime context-aware sanitizer instrumentation. & RXSS, SXSS&
		\begin{minipage} [t] {0.27\textwidth} 
			\begin{itemize}
				\item not suitable for encoded scripts and CSS file scripts. 
				\item relies on the presence of clean pages at the server. 
				\item high false positive and negative rates due to the ignorance of dynamically generated scripts.
			\end{itemize} 
		\end{minipage}\\
		
		\noalign{\smallskip}\hline\noalign{\smallskip}
		Gupta and Gupta~\cite{Gupta2018e} & HTTP traffic monitoring and matching stored sanitized JS variables with unsantized ones generated from HTTP responses. Stored sanitized JS variables are collected through performing a dynamic analysis. & Runtime context-aware sanitizer instrumentation. & WXSS &
		\begin{minipage} [t] {0.27\textwidth} 
			\begin{itemize}
				\item low detection rate (less than 97\%).
			\end{itemize} 
		\end{minipage}\\
		
		\noalign{\smallskip}\hline\noalign{\smallskip}
		Chaudhary et al.~\cite{Chaudhary2019} & HTTP traffic monitoring through matching stored and generated scripts. & Action authentication with Runtime context-aware sanitization. & WXSS &
		\begin{minipage} [t] {0.27\textwidth} 
			\begin{itemize}
				\item lack of performance overhead analysis.
			\end{itemize} 
		\end{minipage}\\
		
		\noalign{\smallskip}\hline
	\end{longtable}

\section{What type of evidences are used to validate solution proposals? (RQ5)}
\label{sec:rq5}
The diversity of solution proposals enabled the use of different validation techniques and data sources. In this section, we describe the different validation techniques used to check the effectiveness of proposed solutions as well as sources used for experimentation. This allowed us to evaluate the quality of studies included in this review and may help researchers to replicate existing studies.

\subsection{Validation techniques}
\begin{figure}[H]
	\centering
	\includegraphics[width=.8\textwidth]{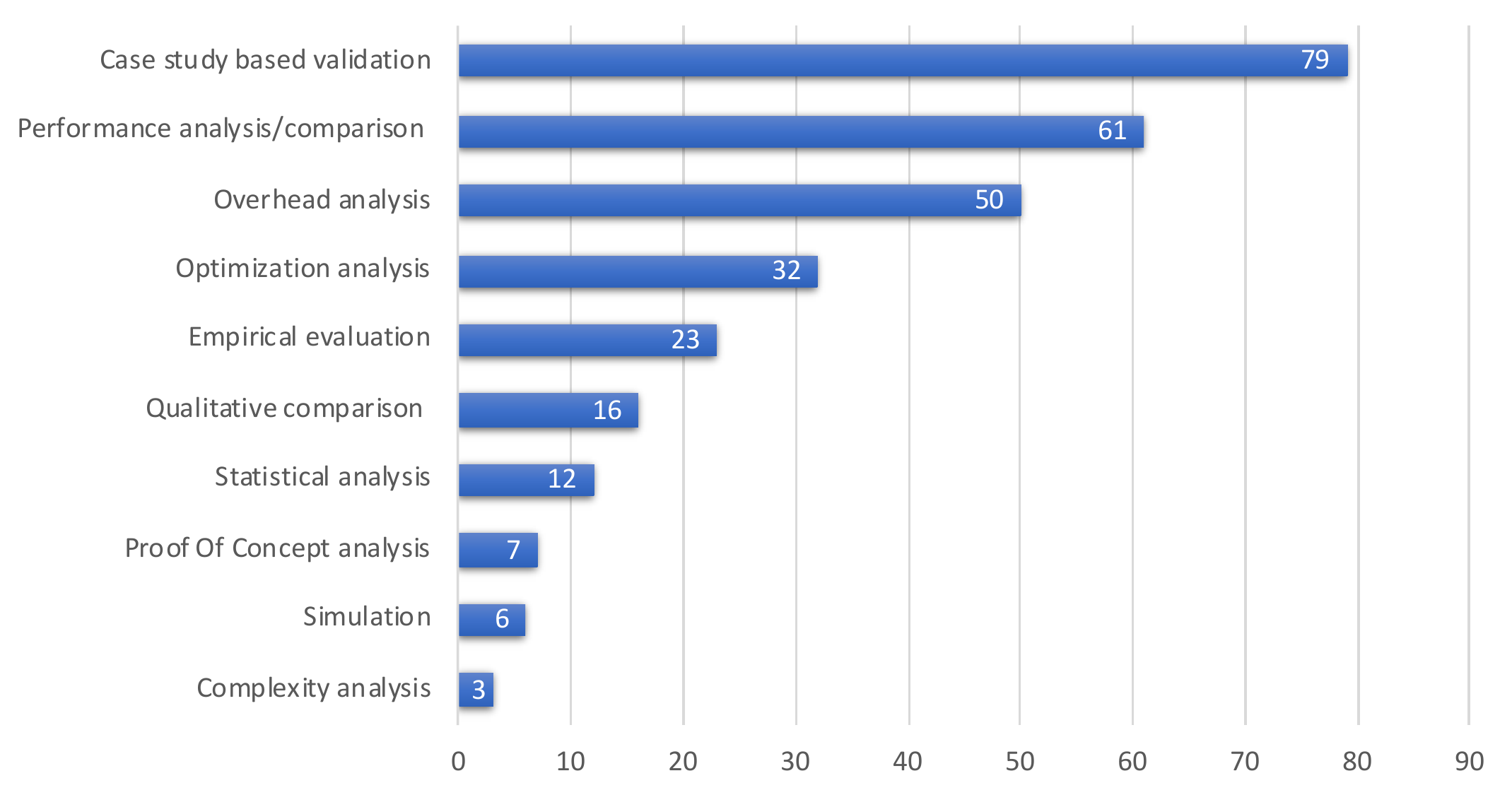}
	\caption{Distribution of validation techniques}
	\label{fig:validation}       
\end{figure}

The adopted data extraction process allows the identification of variant validation techniques adopted by existing studies. Case study, quantitative comparison with similar works and performance overhead analysis are the most adopted techniques as showed in Figure~\ref{fig:validation}.  

\begin{figure}[H]
	\centering
	\includegraphics[width=1\textwidth]{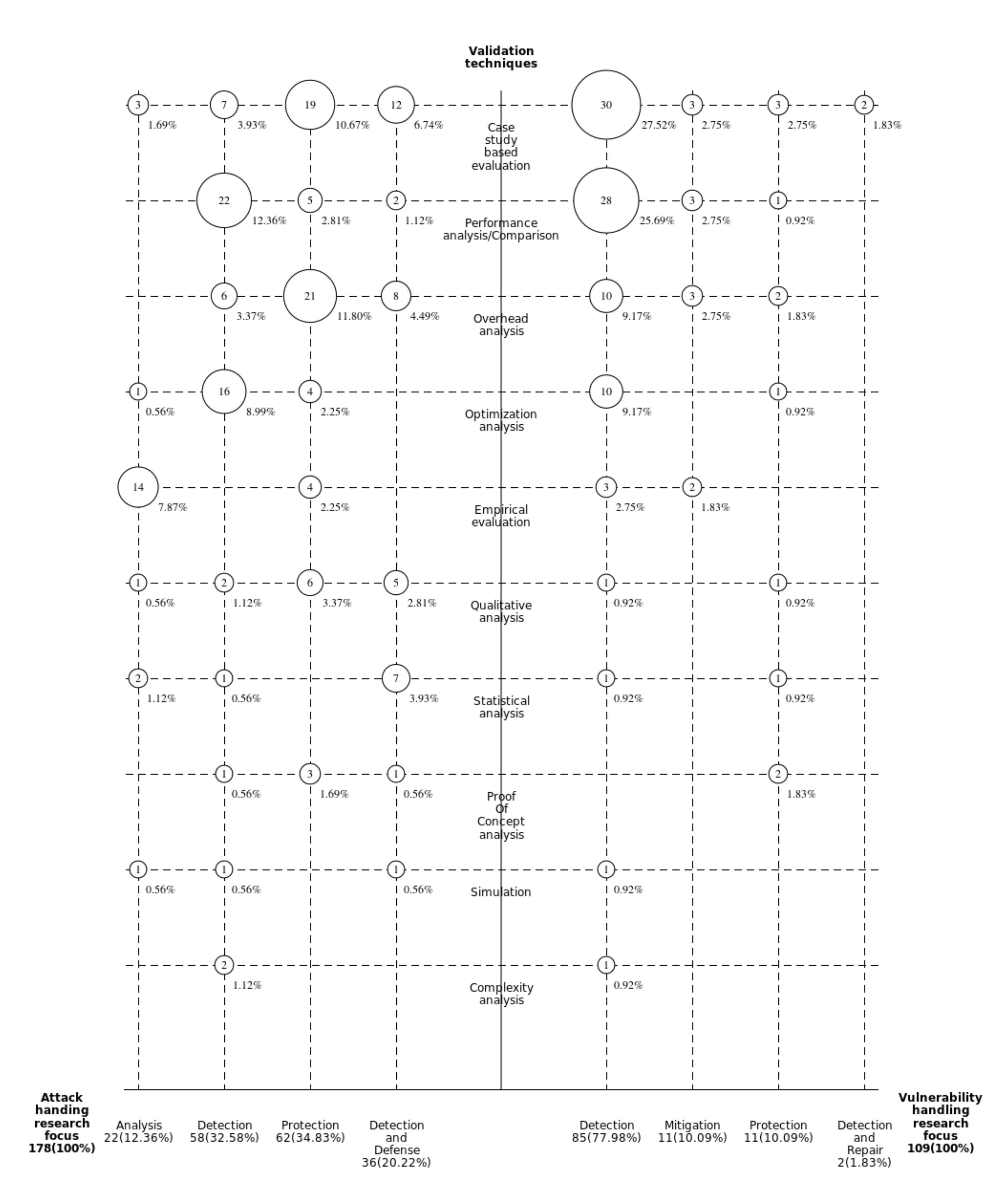}
	\caption{Distribution of validation techniques per contribution to tackle XSS attacks and vulnerabilities}
	\label{fig:validation2}       
\end{figure}

Figure~\ref{fig:validation2} shows refined results where adopted validation techniques are distributed per type of contribution to tackle the attack.
Despite the importance of overhead analysis in runtime detection of attacks, several studies focused only on their detection performances omitting such important factor. This prevents fair comparison of accurate validation of existing proposals and checks their usefulness in practice.

\subsection{Data sources used in validation}
Different publicly available sources for XSS have been used. Those sources are given in Table~\ref{tab:xssvect} together with the list of studies using them. Note that only active links are reported in the table, dead links are ignored. It can be depicted that XSSeed, OWASP and HTML5 Security CheatSheet are the most used sources for retrieving malicious XSS payloads. Alexa top sites are the most experimented URLs and sources to extract benign scripts. Only four complete datasets are made available in Github, Kaggle and figshare. 

	\begin{longtable}{p{2.55cm}p{10cm}lp{1.6cm}}
		\caption{XSS attack vectors data sources: M = Malicious, B = Benign, MB: Mixed}
		\label{tab:xssvect}\\      
		\hline\noalign{\smallskip}
		source & link & type & studies\\
		\noalign{\smallskip}\hline\noalign{\smallskip}
		XSSed project & \url{http://www.xssed.com/} & M & \cite{Nadji2009}, \cite{Bates2010}, \cite{Athanasopoulos2010}, \cite{Nunan2012}, \cite{Pelizzi2012}, \cite{Wang2014}, \cite{Mitropoulos2016}, \cite{Gupta2016}, \cite{Rathore2017}, \cite{Zhang2019}, \cite{Fang2019}, \cite{Kadhim2020}, \cite{Buyukkayhan2020}, \cite{Li2020}, \cite{Malviya2021}, \cite{Tariq2021}, \cite{Mokbal2021}, \cite{Wang2022}, \cite{Liu2022}\\
		Open Bug Bounty & \url{https://www.openbugbounty.org/} & M & \cite{Yamazaki2018}, \cite{Buyukkayhan2020}, \cite{Mokbal2021}\\
		DMoz Archive & \url{http://dmoztools.net/} & B & \cite{Nunan2012}, \cite{Wang2014}\\
		ClueWeb+ & \url{https://www.lemurproject.org/} & B & \cite{Scholte2012a}, \cite{Nunan2012}, \cite{Zhang2019}\\
		Alexa & \url{https://www.alexa.com/topsites} & B & \cite{Van-Acker2012}, \cite{Lekies2013}, \cite{Heiderich2013}, \cite{Stock2014}, \cite{Parameshwaran2015}, \cite{Stock2015}, \cite{Lekies2017}, \cite{Rathore2017}, \cite{Melicher2018}, \cite{Steffens2019}, \cite{Malviya2021}, \cite{Bensalim2021}, \cite{Melicher2021}, \cite{Wang2022}\\
		Github & \url{https://github.com/duoergun0729/1book/tree/master/data} & M & \cite{Zhou2019}, \cite{Tariq2021}\\
		& \url{https://github.com/IramTariq/XSS-attack-detection} & MB & \cite{Tariq2021}\\
		& \url{https://github.com/ajinabraham/OWASP-Xenotix-XSS-Exploit-Framework} & M & \cite{Lv2019}\\
		& \url{https://github.com/bsmali4/XSSfork} & M & \cite{Lv2019}\\
		& \url{https://github.com/fuzzdb-project/fuzzdb} & M & \cite{Bozic2015a}, \cite{Lv2019}\\
		& \url{https://codeload.github.com/foospidy/payloads/zip/master} & M & \cite{Lv2019}\\
		& \url{https://github.com/stivalet/PHP-Vulnerability-test-suite} & MB & \cite{Gupta2016c}\\
		& \url{https://github.com/payloadbox/xss-payload-list} & M & \cite{Chaudhary2021}, \cite{Frempong2021}, \cite{Mokbal2022}\\
		& \url{https://github.com/WhiteRabbitc/Wooyun-Email-XSS-Dataset/tree/master/malious-sample} & M & \cite{Fang2020}\\
		& \url{https://github.com/yahoo/webseclab} & M & \cite{Caturano2021}\\ 
		& \url{https://github.com/danielmiessler/SecLists/blob/master/Fuzzing/XSS/XSS-RSNAKE.txt} & M & \cite{Talib2021}, \cite{Garn2021}\\
		& \url{https://gist.github.com/kurobeats/9a613c9ab68914312cbb415134795b45} & M & \cite{Talib2021}\\
		Kaggle &  \url{https://www.kaggle.com/syedsaqlainhussain/cross-site-scripting-xss-dataset-for-deep-learning} & MB & \cite{Chaudhary2021}, \cite{Mokbal2022}\\
		OWASP project & \url{http://www.owasp.org/index.php/XSS_Filter_Evasion_Cheat_Sheet} & M & \cite{Nadji2009}, \cite{Hooimeijer2011}, \cite{Vernotte2014}, \cite{Javed2014a}, \cite{Bozic2015a}, \cite{Fazzini2015}, \cite{Ahmed2016}, \cite{Pan2016b}, \cite{Gupta2016e}, \cite{Gupta2017a}, \cite{Gupta2018d}, \cite{Gupta2018c}, \cite{Gupta2018e}, \cite{Wijayarathna2019}, \cite{Iqbal2019}, \cite{Gupta2020a}, \cite{Xu2020}, \cite{Talib2021}\\ 
		HTML5 Security Cheatsheet & \url{http://html5sec.org/} & M & \cite{Javed2014a}, \cite{Bozic2015a}, \cite{Fazzini2015}, \cite{Gupta2016b1}, \cite{Gupta2016e}, \cite{Gupta2017a}, \cite{Gupta2018d}, \cite{Gupta2018c}, \cite{Gupta2018e}, \cite{Gupta2019a}, \cite{Gupta2020a}, \cite{Xu2020}, \cite{Chaudhary2020}, \cite{Talib2021}, \cite{Garn2021}\\
		Vulnerability Lab & \url{https://www.vulnerability-lab.com/} & M & \cite{Javed2014a}, \cite{Gupta2016b1}, \cite{Gupta2016e}, \cite{Gupta2017a}, \cite{Gupta2018d}, \cite{Gupta2018c}, \cite{Gupta2018e}, \cite{Gupta2019a}, \cite{Gupta2020a}\\
		Burp suite & \url{http://portswigger.net/burp} & M & \cite{Lv2019}, \cite{Garn2021}, \cite{Wang2022}\\
		CWE+ & \url{https://cwe.mitre.org/data/index.html} & M & \cite{Li2020a}\\
		figshare & \url{https://figshare.com/articles/dataset/XSS_dataset1_csv/13046138/4} & MB & \cite{Mokbal2020}\\
		Impact Cyber Trust & \url{https://www.impactcybertrust.org/dataset_view?idDataset=940} & M & \cite{Mereani2021}\\
		PastBin & \url{https://pastebin.com/BdGXfm0D} & M & \cite{Javed2014a}\\
		JS CheatSheet & \url{https:// htmlcheatsheet.com/js/} & B & \cite{Frempong2021}\\
		Chef Secure & \url{https://chefsecure.com/blog/the-12-exploits-of-xss-mas-infographic} & M & \cite{Frempong2021}\\
		Exploit database & \url{https://www.exploit-db.com/} & M & \cite{Niakanlahiji2019}, \cite{Pazos2021}\\
		Information Technology Laboratory & \url{https://nvd.nist.gov/vuln} & M & \cite{Wurzinger2009}\\

		\noalign{\smallskip}\hline
	\end{longtable}

\section{Issues and open challenges}
\label{sec:discussion}
In this section, we discuss limitations, issues and open challenges preventing the proper handling of XSS attacks. Those issues and challenges are derived from the findings of the present review.

\subsection{Lack of appropriate vulnerability repair mechanisms}
Handling XSS vulnerabilities is a first step toward defending against XSS attacks. But vulnerability repair has earned less attention compared with detection and mitigation. Mitigation techniques are important but have several unstudied flaws. Secure APIs require much time for their development, feasibility check, standardization and adoption by different browsers. Secure templating are framework dependent and they are still in their infancy stage. Sanitization is an intuitive solution to eliminate XSS vulnerabilities but more automatic tools and guidelines are need to be provided for their appropriate and easy usage by inexperienced developers. Therefore, automatic and easy to use tools for detection with repair capabilities of XSS vulnerabilities are still an open research direction that has not yet been sufficiently explored.

\subsection{Limitations of vulnerability detection techniques}
For the detection of XSS vulnerabilities, both static and dynamic analysis are explored by existing studies. Those approaches have several known limitations. Static analysis based approaches require the availability of the code of tested web applications and suffers from reporting high false positive rates~\cite{Marashdih2019}. Besides those general limitations, individual static based solutions have other specific flaws. Model driven-based code auditing is unsuitable for large web applications; manual specification of models becomes complicated; besides, load and manipulation of large models is  cost-effective. Machine learning based code auditing suffers from the lack of interpretability of results; the black-box nature of classifiers prevents understanding their classification processes. The use of gray-box classifiers such as J48 used in~\cite{Gupta2016c} may alleviate this issue since paths can be extracted from trained models which enables the understandability of the classification scheme. Static taint-analysis suffers from the inability to handle dynamically generated scripts at browsers which is a common practice that is widely used in advanced web applications. In the other side, dynamic analysis has also some limitations; as the number of injection points increases, it takes longer to detect all XSS vulnerabilities. Moreover, it produces high false negatives due to their reliance on a set of payloads that are often unable to cover all possible paths. Despite the extensive interest in test-case generation, those techniques are still ineffective for the detection of all real-world scenarios specially with the use of obfuscations. Hybrid analysis provides a trade-off between static and dynamic analysis, however only three studies are found in the literature, thereby we advocate more studies in this direction. Moreover, much interest should be given to handle vulnerability flaws across multiple pages; only few studies are addressing such an issue.  

\subsection{Bias of interest toward basic XSS attacks}
Several XSS attack types are reported and explored in the literature. However, the different types are not treated equally. The study revealed a bias toward the three basic attacks (i.e., RXSS, SXSS and DXSS). The other attacks are not sufficiently studied and more attacks may be in the way. This can be justified by the fact that basic attacks are well-recognized by web security communities and are well-documented. The experience showed that no single solution is able to handle all the attack types, hence more appropriate solutions should be explored to defend against other variants of XSS attacks.

\subsection{Limitations of attack detection techniques}
The study revealed that machine learning is the widely explored approach for the detection of XSS attacks. We agree that machine learning is a promising approach for the detection of cybersecurity attacks including XSS. However, three main obstacles hinder their usage in practice: \textit{lack of interpretability of their predictions}, their \textit{susceptibility to adversarial attacks} and \textit{lack of benchmarks}~\cite{Hannousse2021, Hannousse2021a}. The effective adoption of machine learning for cybersecurity attack detection requires to properly mitigate these three problems. In the review, only two studies have been found addressing the interpretability of their proposed models~\cite{Zhou2019, Mereani2021}. In ~\cite{Zhou2019}, Bayes Networks are adopted since they provide clear semantics that enable learning probability distributions from data~\cite{Mihaljevic2021}. In~\cite{Mereani2021}, the authors proposed deriving explainable rules from black-box models that make the predictions generative and explainable. The latter approach is limited to models using binary features which does not fit to all discriminating features explored in the literature. Regarding susceptibility to adversarial attacks, only four studies addressed the issue, reinforcement learning is the most adopted solution to improve the performance of machine learning models through learning and generation of adversarial attack vectors and retrain models on those attacks. More techniques should be explored in this direction. Additionally, benchmarking is a long standing problem for machine learning based proposals. Benchmarks enable fair comparison of existing solutions and the development of more robust models~\cite{Hannousse2021a}. In \cite{Mokbal2020}, the authors cooperate by proposing a new oversampling algorithm for XSS datasets. More researches are also advocated to deal with such problem.

\subsection{Limitations of attack defense techniques}
Due to the rapid development of attack tactics, traditional filtering techniques such as white/blacklist, Regex pattern matching become inefficient for the detection of all XSS attacks. They need to be frequently updated whenever new attack vectors are reported. Surprisingly, despite the prevalence of web applications with file upload capabilities, such as online social networks, upload filtering is getting less attention. Only two studies covered the issue but with unsupported limitations~\cite{Barth2009, Barua2011}. We advocate more studies that explore properly the upload filtering technique. In the other side, content security policies (CSP) is a powerful and robust filtering solution, if used properly, to defend against XSS attacks. The ideal solution requires to deal with scalability, compatibility, dynamic generation of elements and automatic generation of policies. Those issues are partially addressed by contemporary studies. The study of Xu et al.~\cite{Xu2020} is the only contribution targeting the scalability issue of CSP policies where fine-grained rules are enabled by the proposed solution. Fine-grained policies enable the specification of allowed and/or blocked scripts at elementary and specific tags. The automatic generation of CSP policies for web applications is also mandatory for two reasons: (1) avoid potential errors caused by manual specifications and (2) handle complex and large scale web applications. The automatic generation is addressed in~\cite{Doupe2013, Fazzini2015, Pan2016}. However, those solutions have several limitations related to high false positive rates and performance overheads. Therefore, more improvements are required to reach the ideal CSP policy based solution. Moreover, dealing with dynamic element integration is mainly treated by blacklisting the \textit{eval()} function such in~\cite{Xu2020}. The same problem is faced by MTD based techniques. More solutions should be explored such as adopting effective combinations of static and dynamic analysis for the automatic generation of policies. This way, dynamically added elements can be covered by policies as well as randomization mechanisms used by MTD based approaches. There is still an obstacle facing the effectiveness of CSP policies which is their susceptibilities to CR-XSS attacks. Consequently, more studies are needed to explore how to cope with CR-XSS attacks in CSP based policies.  

\subsection{Bias toward JavaScript based XSS}
Cross-site scripting attacks are primarily caused by the execution of injected malicious scripts making benefits of unsanitized input data. Those scripts are not necessary written in JavaScript, other scripting languages such as VBScript, ActiveX, Flash, and even CSS can be used. However, a remarkable bias is shown toward XSS attacks caused by JavaScripts. The bias is understandable because of the popularity of JavaScript. However, for completeness and robustness, other XSS vectors written in other scripting languages should also be taken into account. An initiative  was taken in~\cite{Bisht2008}, by preventing the execution of scripts embedded on Flash objects. Bensalim et al.~\cite{Bensalim2021} also considered attacks that can come from embedding scripts into JSON files. More deep investigations are required to deal with all XSS attack vectors.  
In addition, dealing with HTML5 specific features when designing new defense mechanisms starts getting much attention. Fifteen included studies proposed defense mechanisms that also work with HTML5 embedded attacks~\cite{Gupta2018c, Gupta2020a, Wang2021, Bensalim2021}. Developers of new defense attacks should follow the same path.

\section{Threats to validity}
\label{sec:tv}
This section discusses the validity threats of our conducted study. As suggested by Ampatzoglou et al.~\cite{Ampatzoglou2019}, we discuss three types of threats to validity (study selection, data, research) and we show how we struggled to mitigate them.

\subsection{Study selection validity}
\label{sec:sts}
The identification of relevant papers is the key point of any secondary study. To avoid missing relevant papers, we queried six well-known academic libraries in the field. In addition to those suggested by Kuhrmann et al.~\cite{Kuhrmann2017}, we also searched papers in Scopus. The automatic search process is based on the presence of search keywords in titles only. Since this may affect the results, we completed the search with exhaustive and recursive snowballing considering titles and abstracts. At the end of the search process, a separate automatic search was made with Google Scholar and no additional relevant papers were retrieved. 
After relevance and inclusion/exclusion criteria check, we conducted a strict quality assessment process. This was the cause of the elimination of a great number of studies. However, we believe that our quality assessment criteria were objective. We did not penalize early publications for their number of citations and we used recognized sources for ranking. In addition, each of the other criterion was adopted for a specific purpose as discussed in section~\ref{sec:qa}.
The only limitation that can be reported for study selection is the ignorance of non-academic studies. The problem is caused by the absence of well-recognized sources and the need for different quality assessment criteria for grey literature such as technical reports and feature papers.

\subsection{Study data validity}
\label{sec:dv}
Data extraction from included studies is performed and checked separately by the authors. This enabled the identification of potential inconsistencies. When they appear, the authors performed a second check and conflicts are resolved after discussion. In addition, concise templates are used for extraction, this is designed by the authors at the planning phase before conducting the review.

\subsection{Study research validity}
\label{sec:rv}
The results derived from the present study only concerns high quality published studies in academic repositories. Grey literature, non-English and low quality papers are excluded from the review. Specifically several non-English and low-quality papers were found published in the literature, however, sticking to the adopted protocol, those studies were excluded. 

\section{Conclusion}
\label{sec:conclusion}
In this paper, we conducted a systematic mapping and a comprehensive survey studying the advancement in research to tackle XSS attacks. The study is not restricted to a period of time and covered high quality studies published since its discovery. Several studies were found in the literature  but a remarkable interest is only observed in the last few years. Despite the diversity of solutions proposed over the years, XSS attacks are still prevalent and targeting new web applications and platforms. The study revealed much attention to XSS vulnerability detection instead of its repair. For securing web navigation, several defense lines should be provided. As a staring point, developers need to be aware of the consequences resulted from ignoring security practices, at the same time, effective tools enabling the automatic detection and repair of XSS vulnerabilities should be made available to facilitate their tasks. Dynamic defense techniques against XSS attacks should also be provided for protecting innocent users when new attacks occur, those techniques are still immature for providing the intended protection level against all types of XSS attacks~\cite{Zheng2021}. Traditional filtering approaches become ineffective regarding the new developed web technologies, more advanced techniques need to be explored. Although the wide adoption of machine learning based techniques for the detection of XSS attacks, existing endeavors only focus on performance analysis omitting three important problems related to cyber-security communities (1) interpretability of prediction results, (2) robustness of models against adversarial attacks and (3) suitability for integration in real-world architectures and platforms. The review also denoted a bias toward basic XSS attacks; this needs to be alleviated by advocating more research targeting other XSS attack variants, specifically WXSS and XAS that are targeting online social networks that may affect wider populations in a short period of time. Moreover, regarding the rapid development of web technologies,  XSS attacks written in other scripting languages such VBScript and ActiveX or embedded in advanced web languages such as HTML5 need to properly be studied for completeness and robustness.

\flushend 

\end{document}